\documentclass{jfm}

\usepackage{graphicx}
\usepackage{natbib}
\usepackage{amsmath}
\usepackage{color}

\ifCUPmtlplainloaded \else
  \checkfont{eurm10}
  \iffontfound
    \IfFileExists{upmath.sty}
      {\typeout{^^JFound AMS Euler Roman fonts on the system,
                   using the 'upmath' package.^^J}%
       \usepackage{upmath}}
      {\typeout{^^JFound AMS Euler Roman fonts on the system, but you
                   dont seem to have the}%
       \typeout{'upmath' package installed. JFM.cls can take advantage
                 of these fonts,^^Jif you use 'upmath' package.^^J}%
       \providecommand\upi{\pi}%
      }
  \else
    \providecommand\upi{\pi}%
  \fi
\fi

\ifCUPmtlplainloaded \else
  \checkfont{msam10}
  \iffontfound
    \IfFileExists{amssymb.sty}
      {\typeout{^^JFound AMS Symbol fonts on the system, using the
                'amssymb' package.^^J}%
       \usepackage{amssymb}%
         \let\leq=\leqslant
         \let\geq=\geqslant
      }{}
  \fi
\fi

\ifCUPmtlplainloaded \else
  \IfFileExists{amsbsy.sty}
    {\typeout{^^JFound the 'amsbsy' package on the system, using it.^^J}%
     \usepackage{amsbsy}}
    {\providecommand\boldsymbol[1]{\mbox{\boldmath $##1$}}}
\fi

\providecommand\bnabla{\boldsymbol{\nabla}}
\providecommand\bcdot{\boldsymbol{\cdot}}
\newcommand\bfbeta{\boldsymbol{\beta}}
\newcommand\bfeta{\boldsymbol{\eta}}
\newcommand\bfxi{\boldsymbol{\xi}}
\newcommand\bfzero{\boldsymbol{0}}
\newcommand\bfb{\boldsymbol{b}}
\newcommand\bfc{\boldsymbol{c}}
\newcommand\bfd{\boldsymbol{d}}
\newcommand\bfh{\boldsymbol{h}}
\newcommand\bfm{\boldsymbol{m}}
\newcommand\bfu{\boldsymbol{u}}
\newcommand\bfw{\boldsymbol{w}}
\newcommand\bfx{\boldsymbol{x}}
\newcommand\bfy{\boldsymbol{y}}
\newcommand\bfz{\boldsymbol{z}}
\newcommand\Rey{\mbox{\textit{Re}}}  
\newcommand\bfLambda{\mathsfbi{\Lambda}}
\newcommand\bszero{\mathsfbi{0}}
\newcommand\bfA{\mathsfbi{A}}
\newcommand\bfB{\mathsfbi{B}}
\newcommand\bfL{\mathsfbi{L}}
\newcommand\bfQ{\mathsfbi{Q}}
\newcommand\bfT{\mathsfbi{T}}

\newtheorem{theorem}{Theorem}

\title[On long-term boundedness of Galerkin models]
{On long-term boundedness of Galerkin models}

\author[M.~Schlegel and\ B.~R.~Noack]
{Michael Schlegel$^{1,2}$%
\thanks{Author to whom correspondence should be addressed: michael.schlegel@tu-berlin.de}\ns
and Bernd R. Noack$^1$}

\affiliation{
$^1$ Institute PPRIME, CNRS -- Universit\'e de Poitiers -- ENSMA, UPR 3346,
D\'epartement Fluides, Thermique, Combustion, CEAT, 43, rue de l'A\'erodrome,
F-86036 Poitiers cedex, France\\[\affilskip]
$^2$ Institut f\"ur Str\"omungsmechanik und Technische
Akustik, Technische Universit\"at Berlin~MB1,
Stra{\ss}e des 17.\ Juni 135, D-10623 Berlin, Germany
}

\begin{document}

\maketitle

\begin{abstract}
We investigate linear-quadratic dynamical systems with energy preserving qua\-dra\-tic terms.
These systems arise for instance as Galerkin
systems of incompressible flows. A criterion is presented to ensure long-term
boundedness of the system dynamics. If the criterion is violated, 
a globally stable attractor cannot exist for an effective nonlinearity.
Thus, the criterion represents
a minimum requirement for a physically justified Galerkin model of viscous fluid flows. 
The criterion is exemplified e.g. for Galerkin systems of
two-dimensional cylinder wake flow models in the transient and the post-transient regime,
the Lorenz system,
and for physical design of a Trefethen-Reddy Galerkin system. There are
numerous potential applications of the criterion, for instance system reduction and
control of strongly nonlinear dynamical systems.
\end{abstract}

\section{Introduction}
\label{sec:intro}

Focus of this paper is the a priori characterisation 
of the long-term behaviour of a linear-quadratic differential equation system
with energy preserving quadratic term.
Such a dynamical system can be obtained by the spectral discretisation 
of the Navier-Stokes equation.
More generally, many traditional Galerkin models
with orthonormal basis functions fall in this category
\citep{Flechter1984book}.
Of particular interest is the long-term behaviour and attractor properties
which can be ideally extracted analytically from the dynamical system.
For instance, a meaningful model can be requested to have globally bounded solutions.
Respective analytical methods for linear-quadratic Galerkin systems 
are still in their infancy.
For a variety of related problems, 
e.g., properties of fixed points, 
efficient tools for dynamical system analyses
\citep[see, e.g.,][]{Guckenheimer1986book, Khalil2002book} 
and tensor structure analyses have been well elaborated
\citep[see, e.g.,][]{Kolda2009siam}. 

In this study, focus will be placed on 
low-order Galerkin models of the coherent flow dynamics
as a simple starting point.
These models are of particular interest for the understanding
of the nonlinear dynamics \citep[see, e.g.,][]{Holmes2012book}
and are key enablers of closed-loop flow control applications 
\citep[see, e.g.,][]{Noack2011book}.
Examples of low-order models
include boundary layers \citep{Rempfer1994jfm},
cylinder wakes \citep{Deane1991pfa,Noack2003jfm},
mixing layers \citep{Noack2005jfm,Wei2009jfm},
lid-driven cavities \citep{Cazemier1998pf,Balajewicz2013jfm}, and
supersonic diffuser flows \citep{Willcox2005siamjsc}.
However, these models tend to be fragile:
Small changes of system parameters may give rise to unphysical divergent solutions,
at least for a subset of initial conditions.
Thus, parameter identification is a delicate task 
and  a priori knowledge about 
the long-term behaviour of Galerkin models for all initial conditions
is highly desirable.

For a priori analyses of the long-term behaviour, 
the optimum is represented by ana\-ly\-ti\-cal solutions. 
However, the general analytical solution of
a class of linear-quadratic differential equation systems, 
including, e.g., the Lorenz system, 
appears to be unrealisable within the frame of the current state of the art.
In contrast, the simple analytical structure  of a linear-quadratic
Galerkin system provides a key enabler for the 
application of a rich kaleidoscope of the methodologies provided by the theory of 
nonlinear dynamics and control theory.
One example is given by the utilisation of Lyapunov's direct method \citep{Lyapunov1892book}.
In fluid mechanics this method is adopted mainly for two purposes.
Firstly, it is employed
for nonlinear stability analyses of fixed points and for model-based flow control design.
The methodology is well-established 
for linear systems~\citep[see, e.g.,][]{Kim2007arfm,Sipp2010amr},
and generalised 
for nonlinear systems~\citep[see, e.g.,][]{Aamo2002book,Khalil2002book}. 
Applications for Lyapunov-based flow control design are demonstrated
in numerical and experimental investigations~\citep[see, e.g.,][]{Gerhard2003aiaa,Samimy2007jfm,Schlegel2009springer,Schlegel2012jfm}.
A second purpose of the direct Lyapunov method
is to ensure hydrodynamic stability
via the sufficient condition for a monotonically decreasing fluctuation
energy~\citep[see, e.g.,][]{Joseph1976book,Drazin1981book}.
This leads to the identification of lower bounds for the critical Reynolds number
of laminar-turbulent transition and 
the identification of flow structures of maximal energy growth.

However, the application range of Lyapunov's direct method is restricted 
by the lack of a systematic approach for the construction of appropriate Lyapunov functions.
The usage of conventional Lyapunov functions
like the total kinetic energy might fail the desired purpose, 
e.g. to show stability for interior flows:
The linear stability matrix is far from being normal
over a large range of Reynolds numbers.
Temporal energy growth is observed which is
traced back to interactions of non-orthogonal eigenvectors \citep{Trefethen1993science,Schmid2001book}.
The application range of Lyapunov's direct method is enhanced for some
configurations~\citep{Galdi1990arma,Straughan2004book}. 
However, 
the identification of strict lower bounds of flow stability 
is often far below the critical Reynolds number.

\begin{figure}
\begin{center}
\includegraphics[height=4.7cm]{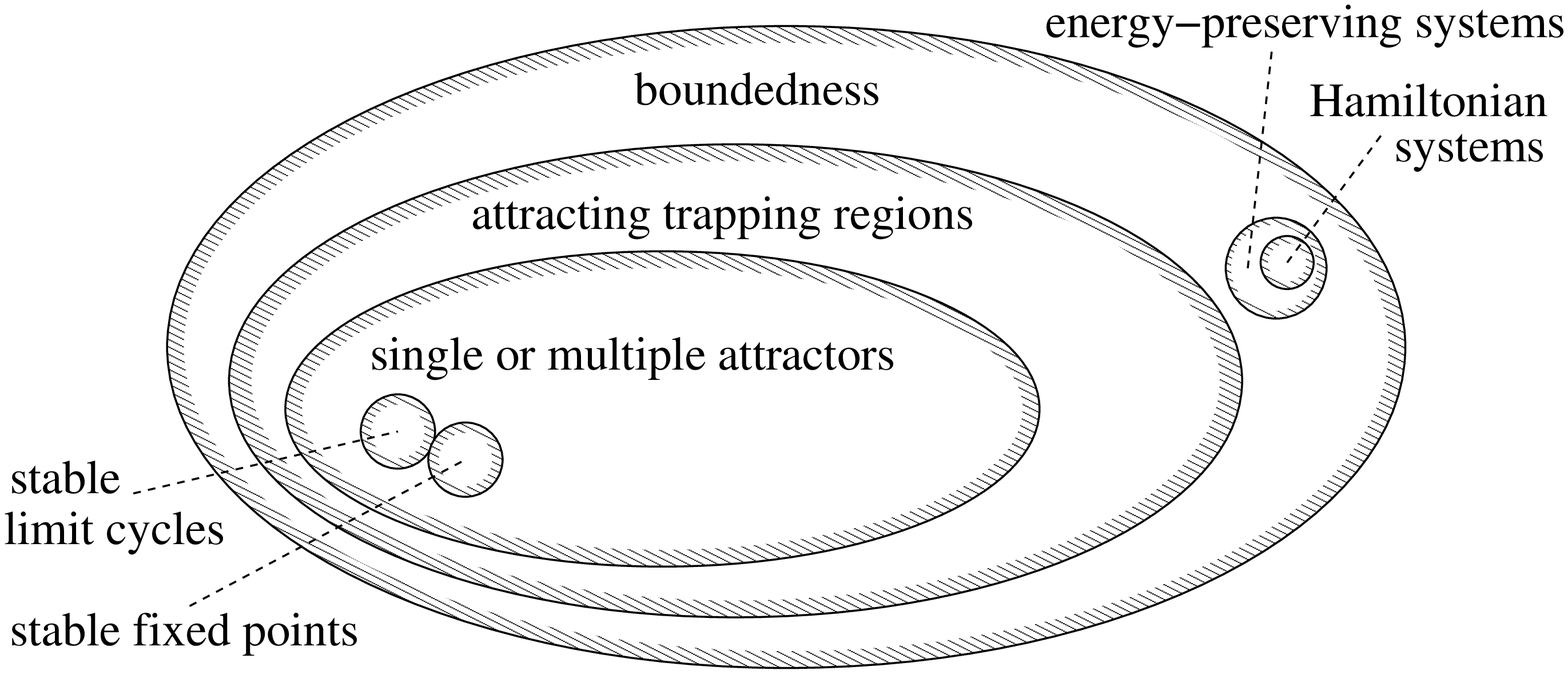}
\caption{Venn diagram of dynamical systems.}
\label{fig:venndiagram}
\end{center}
\end{figure}

The difficulty of finding an appropriate Lyapunov function 
can partially be ascribed by the design goal,
e.g. by ensuring the \textit{global} stability of fixed points
in the \textit{whole} phase space. 
Instead, conditions for the existence of
an arbitrary globally attractive solution are considered. 
In this paper, 
we focus on the existence of trapping regions 
employing Lyapunov's direct method~\citep{SwinnertonDyer2000dynstabsys}.
Coarsely stated, a trapping region is a domain in state space
such that each trajectory once entered the trapping region will remain inside the
trapping region for all times~\citep{Meiss2007book}. 
In case of a (globally) attracting trapping region,
all trajectories outside of the trapping region converge to the trapping region.
The assumption of the existence of an attracting trapping region is connected with
several other properties of dynamical systems 
(see figure~\ref{fig:venndiagram}).
In case that an attracting trapping regions exists, 
the system dynamics is long-term bounded  
because the habitat of the long-term dynamics is represented
by this trapping region. 
An existing (globally stable) single or multiple attractor
must be embedded inside of an attracting trapping region. 

The existence of an attracting trapping region 
can be ensured via the existence of a function
which is strictly of Lyapunov function type 
outside of a trapping region. 
As a well-known example,
the existence of an attracting trapping region is shown
for the Lorenz system~\citep{SwinnertonDyer2001physletta}. 
We follow this hint to investigate long-term boundedness 
and to estimate the extent of existing attractors.
In particular, 
we focus on the generality for the considered class 
of linear-quadratic dynamical systems 
and the simplicity of the construction of the respective Lyapunov functions.

The content of this paper is structured as following:
In section~\ref{sec:models}, the class of the considered dynamical systems
is defined. 
In section~\ref{sec:introductoryexample}, Lyapunov's direct method
is generalised for the identification of trapping regions.
A criterion for long-term boundedness and existence of globally stable attractors
is proposed in section~\ref{sec:stabcrit} and its range of validity is identified.
Analytical and numerical application results are
demonstrated for the investigation of long-term boundedness of
Galerkin systems for the post-transient and transient
dynamics of a two-dimensional cylinder wake flow
in section~\ref{sec:2Dcylindermodels}.
Moreover, a Galerkin system featuring nonlinear characteristics of the Trefethen-Reddy system
is identified in section~\ref{sec:trefethenreddy}.
The Lorenz system is investigated in section~\ref{sec:lorenzsystem}.
In the first appendix section~\ref{sec:preservation}, the generality of
flow configurations is demonstrated, for which the Galerkin method extracts dynamical systems of
the considered class.
In section \ref{sec:conclusions}, the main findings are summarised
and future directions are indicated.

\section{Galerkin models of fluid flows}
\label{sec:models}

In this section, 
the considered class of dynamical systems is introduced.
These systems naturally arise as Galerkin models  
of the incompressible Navier-Stokes equation in a steady domain $\Omega$
with stationary boundary conditions~\citep[see, e.g.][]{Holmes2012book}. 
Galerkin models are typically extracted in two steps.
First, a finite-dimensional Hilbert function subspace~${\mathcal H}$ is chosen.
This subspace is spanned by space dependent modes ${\bfu}_i$, $i=1\ldots N$,
which form an orthonormal basis in this Hilbert space.
The flow ${\bfu}$ is modelled by a Galerkin approximation
with a base flow ${\bfu}_0$ 
and an expansion for the  fluctuation ${\bfu}^\prime={\bfu} - {\bfu}_0$:
\begin{eqnarray}
{\bfu}({\bfxi},t) & = & {\bfu}_0({\bfxi}) \, + \, \sum_{i=1}^N x_i(t) \, {\bfu}_i({\bfxi}). 
\label{eq:galerkinapproximation}
\end{eqnarray}
The flow state is described by the time-dependent modal amplitudes $x_i$.
The spatial variables are denoted by~${\bfxi}$ and the time by~$t$. 
The base flow~${\bfu}_0$ might represent a steady Navier-Stokes solution or mean flow.
The main purpose for the introduction of the base flow is 
that \eqref{eq:galerkinapproximation}
satisfies the boundary conditions 
for arbitrary choices of modal coefficients $x_i$
\citep{Ladyshenskaya1963book}.
The expansion modes ${\bfu}_i$, $i=1,\ldots,N$, may 
arise from a proper orthogonal decomposition of snapshot data
or from other mathematical considerations \cite{Noack1994jfm}.

In the second step, 
a dynamical system is identified. 
At first, the Navier-Stokes equation is projected 
onto the Hilbert subspace~${\mathcal H}$  \citep[see, e.g.][]{Noack2011book}.
As result of the modelling process,
a class of dynamical systems is considered, 
formulated in the vector space of the
state variable~${\bfx}=[x_1,\ldots,x_N]^\top$ by
\begin{eqnarray}
\frac{dx_i}{dt} & = & c_i + \sum\limits_{j=1}^{N} l_{ij} \, x_j
+ \sum\limits_{j,k=1}^{N} q_{ijk} \, x_j \, x_k
\label{eq:gspre}
\end{eqnarray}
with the real numbers $c_i$, $l_{ij}$, $q_{ijk}$, for $i,j,k=1,\ldots,N$. 
Without loss of generality, the
$q_{ijk}$'s are assumed to be symmetric in the last two indices, i.e.
\begin{eqnarray}
q_{ijk} = q_{ikj}, & \; & i,j,k=1,\ldots,N .
\label{eq:qijksymmetry}
\end{eqnarray}

The quadratic term of \eqref{eq:gspre} 
can be shown to be energy-preserving
for a large class of boundary conditions 
(see appendix section~\ref{sec:preservation}).
This means that
the sums of the quadratic coefficients over index permutations are zero
\begin{eqnarray}
q_{ijk} + q_{ikj} + q_{jik} + q_{jki} + q_{kij} + q_{kji} & = &
2 \, q_{ijk} + 2 \, q_{jik} + 2 \, q_{kij} \; = \; 0\, ,\nonumber\\
i,j,k & = & 1,\ldots,N\, .
\label{eq:enerypreservation}
\end{eqnarray}
This property is postulated for the class of dynamical systems discussed in this paper.

The energy-preserving quadratic term has an important effect 
on the evolution of the fluctuation energy 
\begin{eqnarray}
K & := & \frac 1 2 \sum_{i=1}^N x_i^2 \geq 0.
\label{eq:defdistance}
\end{eqnarray}
If ${\bfu}_0$ is the mean flow, 
then $K$ denotes the standard turbulent kinetic energy (TKE) of statistical fluid mechanics.
The time derivative of $K$  reads 
\begin{eqnarray}
\frac{dK}{dt} \; = \; [\bnabla_{\bfx} K]^\top \, \frac{d{\bfx}}{dt} \; = \; \sum_{i=1}^N x_i \, f_i({\bfx}) \; = \; \sum\limits_{i=1}^N c_i \, x_i \; + \sum\limits_{i,j=1}^N l_{ij} \, x_i \, x_j ,
\label{eq:energygrowth1}
\end{eqnarray}
i.e.\ the quadratic terms cancel each other out by \eqref{eq:enerypreservation}.

Defining the vector ${\bfc}:=\left[c_1,\ldots,c_N\right]^\top$ and 
matrix ${\bfL}:=\left[l_{ij}\right]_{i,j=1}^N$,
the evolution of $K$ is given in a simple
vector-matrix notation via
\begin{eqnarray}
\frac{dK}{dt} \; = \; {\bfc}^\top {\bfx} \; + \; {\bfx}^\top \bfL_S \, {\bfx} \, ,
\label{eq:energygrowth2}
\end{eqnarray}
where the symmetric part $\bfL_S:=\,\left( {\bfL} + {\bfL}^\top \right) / 2$ of ${\bfL}$ is introduced.

\section{A generalisation of Lyapunov's direct method}
\label{sec:introductoryexample}

The purpose of this section is twofold.
First, 
a generalised Lyapunov's direct method 
is used 
to guarantee long-term boundedness of a example dynamical system
without the existence of a globally stable fixed point.
Second, an amplitude limiting mechanism  of nonlinear dynamical systems
is elaborated qualitatively and quantitatively, 
requiring the introduction of 'monotonically attracting trapping regions'.
The long-term behaviour is characterised 
by the evolution of the energy
of the system state with respect to the origin.
For mathematical convenience and physical intuition, 
the energy $K$ defined in~\eqref{eq:defdistance} is chosen.
This is modulo factor $1/2$ the square of the
Euclidean distance to the origin given by the Euclidean norm $\Vert.\Vert=\sqrt{2\,K(.)}$.

\subsection{Long-term boundedness of an example system}

\begin{figure}
\includegraphics[height=3cm]{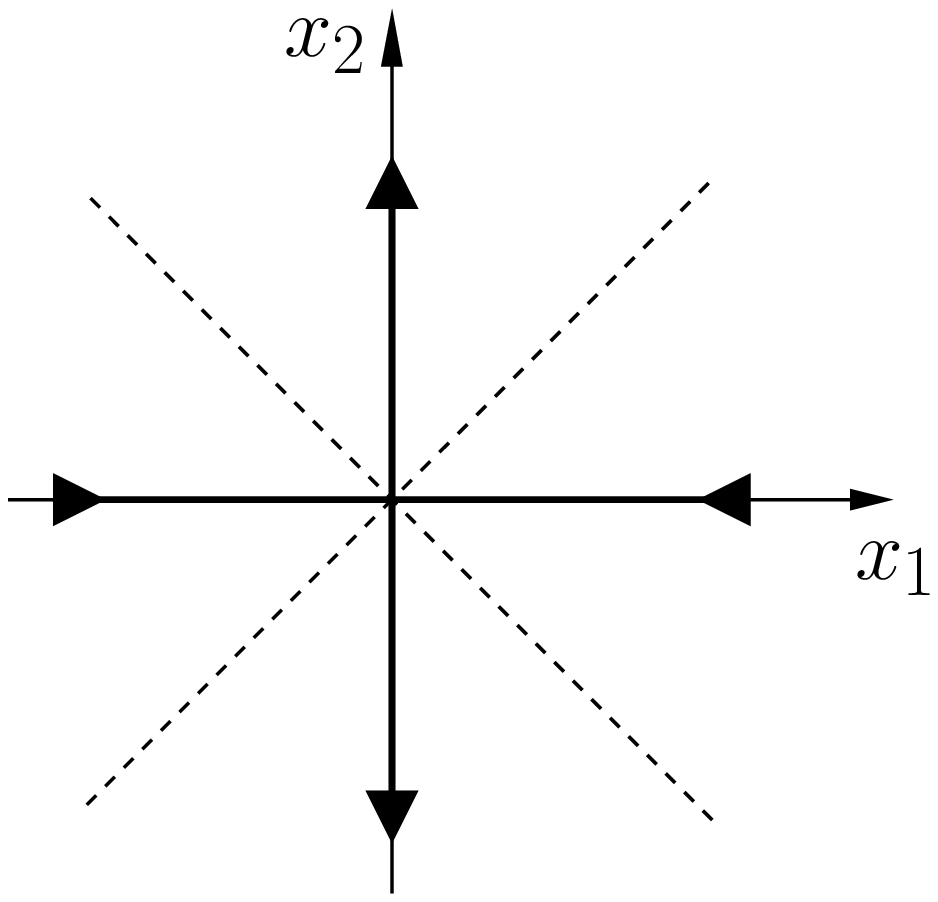}
\hspace{.4cm} \raisebox{1cm}{\Huge +} \hspace{.4cm}
\includegraphics[height=3cm]{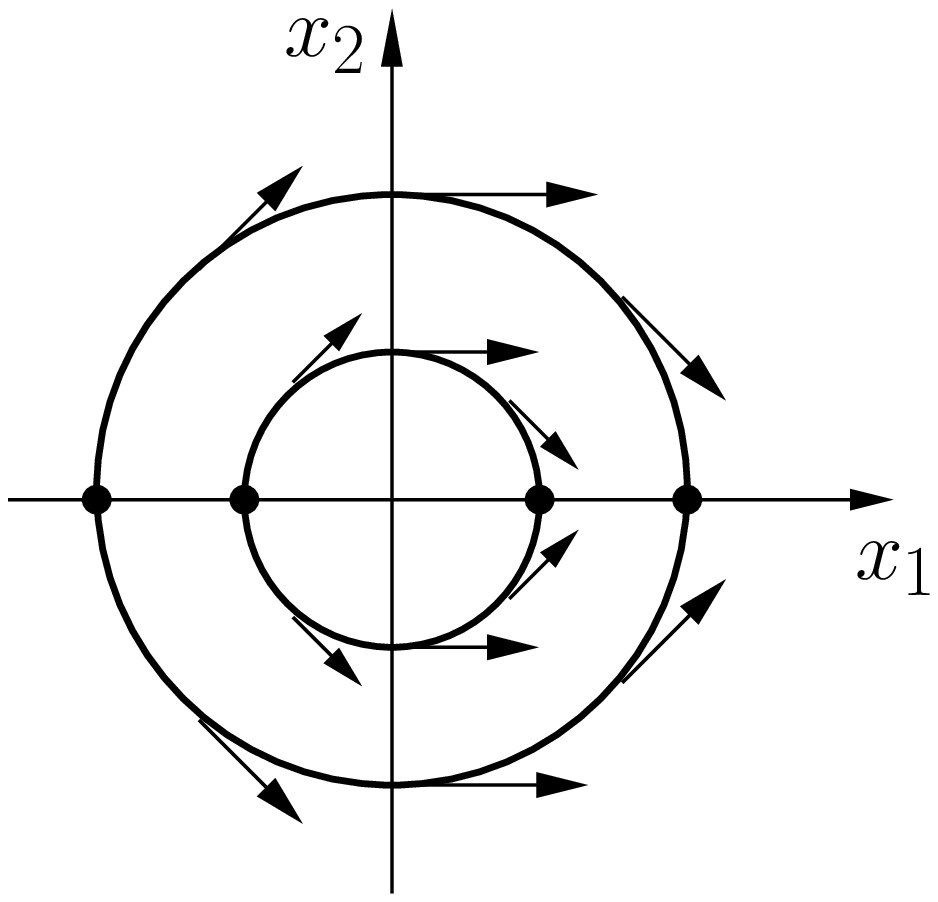}
\hspace{.4cm} \raisebox{1cm}{\Huge =} \hspace{.4cm}
\includegraphics[height=3cm]{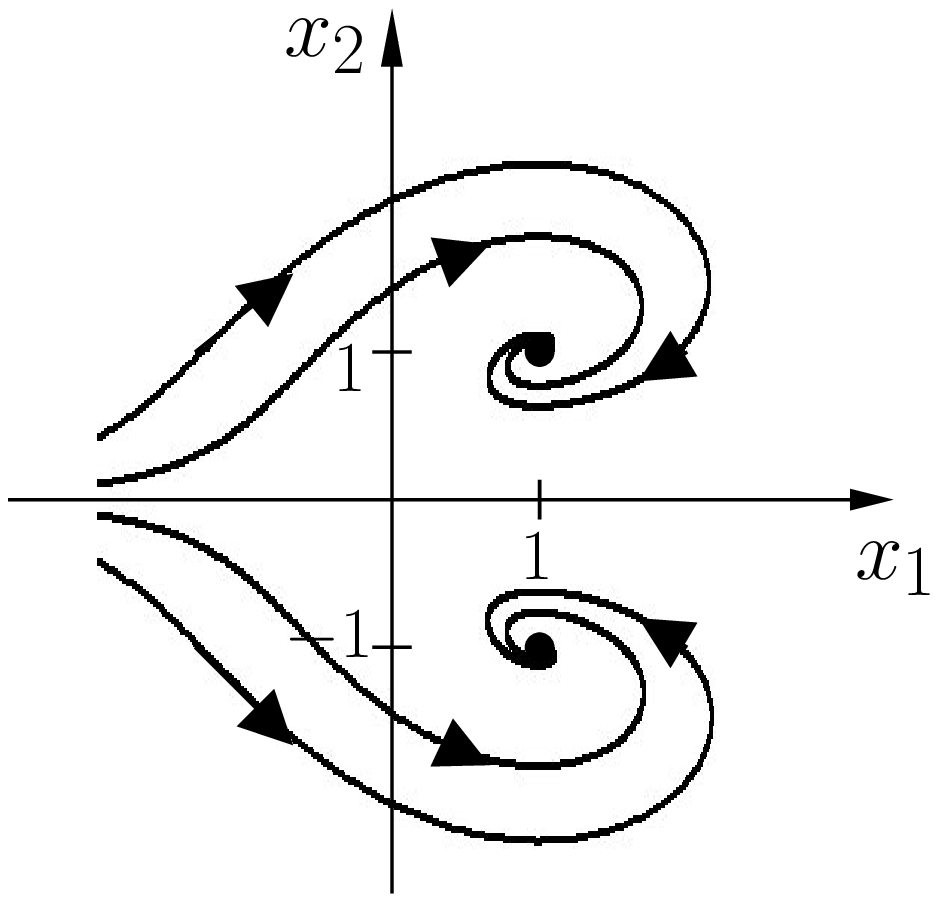}
\caption{Fields of the linear term (left), the quadratic term (middle),
and stable fixed point behaviour (right) of system~\eqref{eq:simplesystem}.}
\label{fig:2Ddynamics}
\end{figure}

As a first example of an energy limiting mechanism, 
the two-dimensional system
\begin{subequations}
\label{eq:simplesystem}
\begin{eqnarray}
\frac{dx_1}{dt} & = & - x_1 \; + \; x_2^2\, ,\\
\frac{dx_2}{dt} & = &  x_2 \; - \; x_1 \, x_2\, ,
\end{eqnarray}
\end{subequations}
is considered.
By the linear, symmetric part~${\bfL}_S={\bfL}=\left[ \begin{array}{cc} -1 & 0\\ 0 & 1 \end{array}\right]$
of the two-dimensional system,
the direction $[0,1]^\top$ with positive energy growth
and the direction $[1,0]^\top$ with negative energy growth are obtained (see figure~\ref{fig:2Ddynamics}).
The field of the quadratic term deflects the trajectories from directions of growing energy
into the directions of shrinking energy.
In result of the interaction of the linear and the quadratic term,
all trajectories are attracted by one of the stable fixed points $[1,1]^\top$, $[1,-1]^\top$,
or along the abscissa to the origin which represents an unstable fixed point.
There is no quadratic Lyapunov function by which the convergence to one of the stable fixed points
can be proven a priori. 
The Lyapunov function does not even exist in the corresponding open
half-planes of attraction.
The energy is increasing or decreasing in dependence of the location of the state.
There are phase-space areas of positive or negative energy growth.
If the dynamics along the trajectory is dominated by negative energy growth,
the system state is attracted e.g.\ to fixed points like in system~\eqref{eq:simplesystem}.
If the dynamics is dominated by positive energy growth, the trajectories may diverge to infinity.

By the transformation
\begin{eqnarray}
{\bfy} & = & {\bfx} - {\bfm}
\label{eq:coordinateshift}
\end{eqnarray}
an arbitrary state ${\bfm}=[m_1,\ldots,m_N]^{\top}$ can be shifted
into the origin of~${\bfy}$. 
The fluctuation energy with respect to~$\bfm$ is defined by
\begin{eqnarray}
K_{\bfm}&:=&\sum_{i=1}^N y_i^2 \; = \; \sum_{i=1}^N (x_i - m_i)^2\, .
\end{eqnarray}
Its evolution is given by
\begin{eqnarray}
\frac d {dt} K_{\bfm} & = & {\bfy}^\top {\bfA}_S \, {\bfy} + {\bfd}^\top {\bfy} \; = \; 
\left( {\bfx} - {\bfm} \right)^\top {\bfA}_S \left( {\bfx} - {\bfm} \right) +
{\bfd}^\top \left({\bfx} - {\bfm}\right) \, ,
\label{eq:energyshift}
\end{eqnarray}
where ${\bfd}$ and ${\bfA}_S$ denote the constant and linear symmetric part of the
transformed Galerkin system.

Employing ${\bfm}=[2,0]^\top$, the example system~\eqref{eq:simplesystem}
is given for the shifted coordinates by
\begin{subequations}
\label{eq:shiftedsimplesystem}
\begin{eqnarray}
\frac{dy_1}{dt} & = & -2 \; - \;y_1 \; + \; y_2^2\, ,\\
\frac{dy_2}{dt} & = & -y_2 \; - \; y_1 \, y_2\, , 
\end{eqnarray}
\end{subequations}
leading to the power balance with respect to the new origin,
\begin{eqnarray}
\frac d {dt} K_{\bfm} & = & -2\, y_1 \; - \; y_1^2 - y_2^2 \, .
\end{eqnarray}
Hence, the energy is growing only in a bounded domain defined by the interior of the
circle given by $(y_1+1)^2 + y_2^2 < 1$.
If $K_{\bfm}$ is large, it will decrease and the trajectories cannot escape each circle~$B$
with the origin at the centre in which
the bounded domain of energy growth is contained.
In conclusion, the long-term dynamics of the shifted system and consequently
of the system~\eqref{eq:simplesystem} are bounded! Outside of each~$B$, the energy~$K_{\bfm}$
represents a strict Lyapunov function and thus Lyapunov's direct method is effective. Inside of~$B$,
the energy~$K_{\bfm}$ can grow and thus Lyapunov's direct method cannot be applied.

\subsection{Monotonically attracting trapping regions}

For a generalisation of this approach, we introduce  monotonically attracting trapping regions.
A \textit{trapping region} $D \subseteq \mathbb{R}^N$ is a compact set, in which each trajectory
remains once it has entered, 
i.e.\ from ${\bfx}(s)\in D$ it follows that 
${\bfx}(t)\in D$ for all $t>s$. 
A trapping region is termed \textit{(globally) monotonically attracting}, 
if an energy is strictly monotonically decreasing 
along all trajectories starting from an arbitrary state outside of $D$. 
This implies that outside of the trapping region the energy
possesses the mathematical properties of a strict Lyapunov function.
For our choice of energy~$K_{\bfm} = \Vert  {\bfx} - {\bfm} \Vert^2/2$, 
it is sufficient to consider closed balls
\begin{eqnarray}
\label{eq:BallX}
B({\bfm},R) & := & \left\{ {\bfx} \in \mathbb{R}^N \, : \, 
\Vert {\bfx} - {\bfm} \Vert^2 \leq R^2 \right\}
\end{eqnarray}
with the centre ${\bfm}$ and radius $0<R<\infty$. 
For later reference, 
these closed balls are also expressed in terms of translated coordinates $\bfy = \bfx -\bfm$:
\begin{eqnarray}
\label{eq:BallY}
B_{\bfy}(R) & := & \left\{ {\bfy} \in \mathbb{R}^N \, : \, 
\Vert {\bfy} \Vert \leq R^2 \right\}.
\end{eqnarray}
Here, each closed ball
containing $D$ as a subset is a monotonically attracting trapping region as well.

If all eigenvalues of ${\bfA}_S$ are negative,
i.e.\ $0>\lambda_1\geq\ldots\geq\lambda_N$, 
then Lyapunov's direct method is effective for large deviations. 
This is immediately shown by the energy evolution equation~\eqref{eq:energyshift}. 
In the following, the energy evolution equation is analysed to obtain monotonically attracting
trapping regions.

For that, a negative definite linear symmetric part~${\bfA}_S$ is postulated in the following.
Invoking the diagonalisation
\begin{eqnarray}
\bfA_S \; = \; {\bfT}^\top \, {\bfLambda} \, {\bfT}
\end{eqnarray}
of the symmetric matrix ${\bfA}_S$ with the diagonal eigenvalue matrix  ${\bfLambda}$ 
and the orthogonal matrix ${\bfT}$ comprising the eigenvectors,
a transformation
\begin{eqnarray}
{\bfz} & = & {\bfT} \, {\bfy}
\end{eqnarray}
is defined, preserving the energy $K_{\bfm}$.
Employing this coordinate transformation (rotation + reflections), 
the energy growth is given by
\begin{eqnarray}
\frac{d}{dt} K_{\bfm} \; = \; \sum\limits_{i=1}^N h_i \, z_i \, + \, \lambda_i \, z_i^2
\; = \; \sum\limits_{i=1}^N \lambda_i \, \left( z_i + \, \frac{h_i}2 \right)^2
\, - \,\sum\limits_{i=1}^N \lambda_i \frac{h_i^2}4 \, .
\label{eq:monotonic3}
\end{eqnarray}
with the components $h_i$ of ${\bfh}\, :=\, {\bfd} \, {\bfT}^\top$.

For ${\bfd}={\bfzero}$ and hence ${\bfh}={\bfzero}$,
the energy is a strict Lyapunov function since $\lambda_i<0$, $i=1,\ldots,N$, has been assumed.
In addition, a globally stable
fixed point is situated at the origin.
For ${\bfd} \neq {\bfzero}$ and hence ${\bfh} \neq {\bfzero}$, 
the energy growth can be positive close to the origin.
In more detail,
the sign of the energy growth is changing at the boundary of an ellipsoid $E$ which is defined via 
\begin{eqnarray}
\sum\limits_{i=1}^N \frac 1{\alpha_i^2} \left( z_i + \, \frac{h_i}2 \right)^2 \; = \; 1 & \mbox{ with } &
\alpha_i \; := \; \sqrt{\frac{\sum_{j=1}^N \lambda_j \, h_j^2} {4 \,\lambda_i}} \, .
\label{eq:ellipsoid}
\end{eqnarray}
In the interior of the ellipsoid~$E$, the energy growth is positive with a maximum growth
at the centre $-{\bfh}/2$. Outside of the ellipsoid the energy is decreasing. At the boundary,
the energy stays constant.
Note, that the origin is situated at the boundary of the ellipsoid because it is trivially
solving equation~\eqref{eq:ellipsoid}. The half-axes~$\alpha_i$ are directly proportional
to the Euclidean norm of~${\bfh}$ and hence of~${\bfd}$.

After a finite time, the system state is trapped in the smallest closed ball $B$ with centre at the origin
of the ${\bfy}$-coordinates
such that the ellipsoid~$E$ is contained\footnote{Only in the non-generic case, that there
is a stable fixed point at the intersection of the boundaries of~$B$ and~$E$, it might take infinite time
to enter $B$.}. Hence, the smallest monotonically attracting trapping ball is given by~$B$!

\begin{figure}
\begin{center}
\includegraphics[height=4cm]{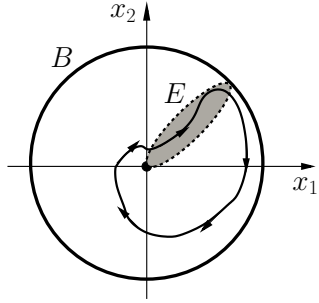}
\caption{Principal sketch of boundedness of a solution of a Galerkin system~\eqref{eq:gspre}
in a closed ball~$B$. For simplicity, the origin is not shifted, i.e.~${\bfm}={\bfzero}$.
The growth of energy to the origin is positive only inside of an ellipsoid $E$.}
\label{fig:ellipsoid}
\end{center}
\end{figure}

The long-term behaviour of the system,
represented e.g. by an globally stable attractor,
is either part of the boundary without growth of~$K_{\bfm}$, or is alternating
between positive energy growth inside of $E$ and negative energy growth
in~$B\setminus E$ (see figure~\ref{fig:ellipsoid}).
The case ${\bfd}={\bfzero}$ can be seen as a degeneration of~$E$ and~$B$
to the fixed point at the origin.

\section{A sufficient criterion for long-term boundedness}
\label{sec:stabcrit}

\begin{figure}
\begin{center}
\includegraphics[height=6.6cm]{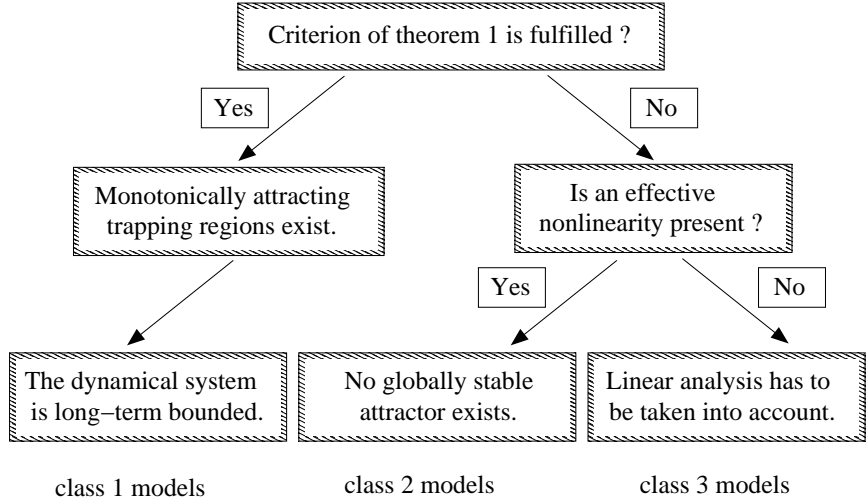}
\caption{Decision tree diagram for long-term boundedness of linear-quadratic systems.}
\label{fig:decisiontree}
\end{center}
\end{figure}

In this section, 
a sufficient criterion for long-term boundedness of Galerkin systems
is derived to exclude infinite blow ups of the system state ${\bfx}(t)$
in finite or infinite periods of time. 
Via the criterion of theorem~\ref{theorem1} below, the existence of
monotonically attracting trapping regions is considered. 
For the generic class of Galerkin
systems~\eqref{eq:gspre} 
with effective nonlinearity, 
it is shown that a globally stable attractor can only exist 
if there is a monotonically attracting trapping region. 
The results of this section
will   culminate in the procedure sketched in figure~\ref{fig:decisiontree},
guiding the determination of the long-term behaviour 
of the Galerkin systems~\eqref{eq:gspre}.

To keep the notation simple, 
the following vector-matrix representation of~\eqref{eq:gspre} is
employed using the symmetric matrices~${\bfQ^{(\alpha)}}:= \left[q_{\alpha ij}\right]_{i,j=1}^N$,~$\alpha=1,\ldots,N$,
\begin{eqnarray}
\frac{d{\bfx}}{dt} & = & {\bfc} + {\bfL}\,{\bfx} +
\left[ \, {\bfx}^\top \, {\bfQ}^{(1)} \, {\bfx} \, ,
\ldots, \, {\bfx}^\top \, {\bfQ}^{(N)} \, {\bfx} \, \right]^\top.
\label{eq:gs}
\end{eqnarray}
Using this nomenclature,
the condition~\eqref{eq:enerypreservation} can be rewritten as
\begin{eqnarray}
q^{(i)}_{jk} + q^{(j)}_{ik} + q^{(k)}_{ij}  = 0, & \;
& i,j,k=1,\ldots,N
\label{eq:enerypreservation2}
\end{eqnarray}
employing the elements $q^{(i)}_{jk}$, $j,k=1,\ldots,N$ of the symmetric matrices ${\bfQ}^{(i)}$.

For the translated variable $\bfy = \bfx - \bfm$, 
the dynamical system
\begin{eqnarray}
\frac{d{\bfy}}{dt} & = & {\bfd} + {\bfA}\,{\bfy}
+ \left[ \, {\bfy}^\top \, {\bfQ}^{(1)} \, {\bfy} \, ,
\ldots, \, {\bfy}^\top \, {\bfQ}^{(N)} \, {\bfy} \, \right]^\top
\label{eq:gsshift}
\end{eqnarray}
with
\begin{eqnarray}
{\bfd}:=\left(c_i + \sum_{j=1}^N l_{ij} m_j + \sum_{j,k=1}^N q_{ijk} m_j m_k \right)_{i=1}^N
\end{eqnarray}
and
\begin{eqnarray}
{\bfA}:= \left( l_{ij} + \sum_{k=1}^N ( q_{ijk} + q_{ikj} ) \, m_k \right)
\end{eqnarray}
is obtained.
Note that the symmetric part $\bfA_S$ of ${\bfA}$ can be represented as
a linear combination of the symmetric part $\bfL_S$ of ${\bfL}$ and of the~${\bfQ}^{(i)}$'s
\begin{eqnarray}
\bfA_S \; := \; \frac 1 2 \left( {\bfA} + {\bfA}^\top \right) & = &
\bfL_S \; - \; \sum\limits_{i=1}^N \, {m_i} \; {\bfQ}^{(i)} \, ,
\label{eq:symmlin}
\end{eqnarray}
exploiting the symmetry
properties~\eqref{eq:enerypreservation} and~\eqref{eq:qijksymmetry}.

Employing the translation, 
a simple condition for the existence
of a monotonically attracting trapping region can be found. 
If all eigenvalues of ${\bfA}_S$ are negative,
i.e.\ $0>\lambda_1\geq\ldots\geq\lambda_N$, 
then the energy evolution equation is transformed
to equation~\eqref{eq:monotonic3} employing the rotation of the coordinate system to the principal axes.
Hence, the domain of energy growth is identified to be inside of the ellipsoid given
by~\eqref{eq:ellipsoid}. Every ball with the origin ${\bfy}={\bfzero}$ at the centre,
which contains the ellipsoid,
is a monotonically attracting trapping region. 
Consider the estimate   
\begin{eqnarray}
\alpha_i & \leq & \sqrt{\frac{\lambda_N}{4\lambda_1}} \, \Vert{\bfd}\Vert\, ,
\end{eqnarray}
invoking the definition~\eqref{eq:ellipsoid} of the half-axes $\alpha_i$.
Then a radius of such a ball,
not necessarily the infinimum amongst such radii, 
is given by $R_m=\sqrt{\lambda_N/\lambda_1} \Vert{\bfd}\Vert$. 

On the other hand, 
given a monotonically attracting trapping region, the distance of every state~${\bfx}$
outside the trapping region to a state~${\bfm}$ inside of the trapping region
is monotonically decreasing by definition.
This is only true if the right-hand side of the power balance~\eqref{eq:energyshift}
is negative for an~${\bfm}$
inside the trapping region and if the energy~$K_{\bfm}$ is large enough.
By symmetry considerations this requires that all eigenvalues of~${\bfA}_S$ are negative.

The above-mentioned results are summarised in the following theorem
\begin{theorem} MONOTONICALLY ATTRACTING TRAPPING REGIONS \\
Regarding the system~\eqref{eq:gs}, the following two statements are equivalent:
\begin{enumerate}
\item There is a monotonically attracting trapping region.
\item There is an ${\bfm}$, such that there are only
negative eigenvalues~$0>\lambda_1\geq\ldots\geq\lambda_N$ of the symmetric linear
part~${\bfA}_S$~\eqref{eq:symmlin} of the shift-transformed system~\eqref{eq:gsshift}.
\end{enumerate}
If these conditions are true, $R_m=\sqrt{\lambda_N/\lambda_1} \>\Vert{\bfd}\Vert$ is a radius
such that $B({\bfm},R_m)$ is a monotonically attracting trapping region.
\label{theorem1}
\end{theorem}

In conclusion, a sufficient condition for the long-term boundedness of systems~\eqref{eq:gs}
over an infinite time horizon has been found.

As a first example of an application of the criterion,
we show 
the existence of a monotonically attracting trapping region
for the system~\eqref{eq:simplesystem}. It
can be written in the form of equation~\eqref{eq:gs} with
\begin{eqnarray}
{\bfc}=\left[ \begin{array}{c} 0\\ 0 \end{array}\right], \;
{\bfL} \; = \; \left[ \begin{array}{cc} -1 & 0\\ 0 & 1 \end{array}\right], \;
{\bfQ^{(1)}} \; = \; \left[ \begin{array}{cc} 0 & 0 \\ 0 & 1  \end{array}\right], \;
{\bfQ^{(2)}} \; = \; \left[ \begin{array}{cc} 0 & -\frac 1 2\\ - \frac 1 2 & 0 \end{array}\right].
\end{eqnarray}
Considering the translation with ${\bfm}=[2,0]^\top$,
and invoking~\eqref{eq:symmlin}, the symmetric part of the transformed system is
\begin{eqnarray}
{\bfA}_S & = &
\left[ \begin{array}{cc} -1 & 0\\ 0 & -1  \end{array}\right].\nonumber 
\end{eqnarray}
By the  negative definiteness of~${\bfA}_S$,
the existence of a monotonically attracting trapping region is shown.
One of these regions is given by the closed ball $B_{\bfy}(2)$ 
in the ${\bfy}$-coordinates (see \eqref{eq:BallY})
and equivalently by $B({\bfm},2)$  in the ${\bfx}$-coordinates (see \eqref{eq:BallX}).
All solutions, given by the stable and unstable fixed points are situated inside of $B$ and
at the boundary of the ellipsoid~$E$ defined by equation~\eqref{eq:ellipsoid},
along which the energy~$K_{\bfm}$ is maintained.

Although boundedness of a large class of systems
is ensured, there might exist long-term bounded systems even with a globally stable fixed point 
which do not fit the condition of theorem~\ref{theorem1}.
One class is given by  Hamiltonian systems in which the  energy~$K_{\bfm}$
is preserved. 
Here, ${\bfA}_S={\bszero}$, because
the distance to an ${\bfm}$ of the initial values remains constant for all times.
All trajectories are embedded in invariant spaces, representing shells of constant distance around the centre.
Hence the dynamics are bounded, but globally stable trapping regions
or even a globally stable attractor do not exist.  

Another class is given from the linear system behaviour of~\eqref{eq:gs} 
with vanishing  nonlinear
term~${\bfQ}^{(i)}={\bszero}$, $i=1,\ldots,N$, ${\bfc}={\bfzero}$, and a non-normal matrix ${\bfL}$.
Here, an attracting behaviour can be accompanied with temporal energy growth as known
from the theory of non-normal matrices \citep{Trefethen2005book}. Even if there is a
globally stable fixed point, the convergence of the trajectories can be non-monotonous
with alternating increase and decrease of the distance to the fixed point. 

However in the generic case of an effective nonlinearity, each globally non-monotonically
attracting set is embedded in a larger monotonically attracting trapping region.
To introduce the notion of an \textit{effective nonlinearity}, a reformulation of the
transformed system~\eqref{eq:gsshift} is considered.
Let ${r}:=\Vert{\bfy}\Vert>0$ be the amplitude (radius) of the state
and ${\bfw}:={\bfy}/{r}$ the `generalised phase' (direction) on the unit ball
$\partial B_{\bfy}(1)$. It yields
\begin{subequations}
\label{eq:gspreall}
\begin{eqnarray}
\frac{d{\bfw}}{dt} & = & \frac 1 r {\bfd} + {\bfA}\,{\bfw} +
{r} \, \left[ \, {\bfw}^\top \, {\bfQ}^{(1)} \, {\bfw} \, , \ldots,
\, {\bfw}^\top \, {\bfQ}^{(N)} \, {\bfw} \, \right]^\top\, ,
\label{eq:gsprephase} \\
\frac{d{r}}{dt} & = & {\bfd}^{\top} {\bfw} +
{r} \; {\bfw}^\top \bfA_S \, {\bfw} \, .
\label{eq:gspreradius}
\end{eqnarray}
\end{subequations}
The focus of interest is on large distances~$r$ from the origin~${\bfm}$.
The nonlinearity is termed effective if
the constant and linear term of the phase equation~\eqref{eq:gsprephase}
can be neglected for large~$r$
\begin{eqnarray}
\frac{d{\bfw}}{dt} & = &
{r} \, \left[ \, {\bfw}^\top \, {\bfQ}^{(1)} \, {\bfw} \, , \ldots,
\, {\bfw}^\top \, {\bfQ}^{(N)} \, {\bfw} \, \right]^\top \, ,
\label{eq:gsphase}
\end{eqnarray}
i.e. the trajectories are driven by the dynamics of the equations~\eqref{eq:gsphase}
and~\eqref{eq:gspreradius}.
Generically, the spatial asymptotical behaviour of the considered Galerkin systems~\eqref{eq:gspre}
is described by these two equations. However, this is not true
if there exist invariant manifolds of~\eqref{eq:gs} with vanishing quadratic term,
i.e.~${\bfx}^\top {\bfQ}^{(i)} {\bfx}={\bfzero}$ for all corresponding states.
Here, linear analyses have to be applied in addition to identify the
long-term behaviour conclusively. For illustration, a corresponding example is detailed in
subsection~\ref{sec:counterexample:Lorenz} of the appendix section~\ref{sec:counterexamples}.
 
For effective nonlinearity,
the dynamics of large deviations is independent from the antisymmetric part of the linear term: 
the dynamics of the phase~${\bfw}$ is determined solely by the quadratic term, the dynamics of~$r$
solely by the symmetric part of the linear term and the constant term. The mechanisms for temporal
energy growth
resulting from non-orthogonal eigendirections of non-normal matrices are excluded for large~$r$.
It can be concluded for each attracting trapping region,
that it is embedded in a monotonically attracting trapping region.

These results are summarised in the following theorem.
\begin{theorem} TRAPPING REGIONS\\ 
Consider a system~\eqref{eq:gs} with effective nonlinearity,
i.e. the dynamics of large deviations of the shifted system~\eqref{eq:gsshift}
can be described by~\eqref{eq:gspreradius} and~\eqref{eq:gsphase}.
Then each (globally) attracting trapping region is contained in a monotonically attracting trapping region.
In particular, the existence of a monotonically attracting trapping region is necessary for the existence
of a globally stable attractor.
\label{theorem2}
\end{theorem}

In conclusion, the long-term behaviour is determined via application of the procedure
illustrated in~figure~\ref{fig:decisiontree}.
For the criterion of theorem~\ref{theorem1},
$N$ components of ${\bfm}$ have to be found such that the $N$ eigenvalues of~${\bfA}_S$
given by~\eqref{eq:symmlin} are negative. While this can be analytically done for systems of
dimension lower than or equal to four, for larger dimensions numerical linear algebra and
multidimensional optimisation like simulated annealing has to be employed.

\section{Long-term boundedness of Galerkin models for cylinder wake flows}
\label{sec:2Dcylindermodels}

In this section, 
we investigate the long-term boundedness 
of a hierarchy of Galerkin models 
for periodic cylinder wakes
\citep{Noack2003jfm}.
The considered systems include  
a 3-dimensional mean-field system (subsection~\ref{subsec:meanfieldsystem}),
an 8-dimensional POD model (subsection~\ref{subsec:8modemodel})
and a 9-dimensional generalisation of this POD model
with a stabilising shift mode 
(subsection \ref{subsec:2Dwakewithshiftmode}).
The existence of a monotonically attracting trapping region is demonstrated 
analytically for the mean-field system 
which is known to have a globally stable limit cycle.
The corresponding existence is also numerically shown 
for the 9-dimensional model 
which generalises the mean-field system
by inclusion of the 2nd to 4th harmonics. 
The existence of a  monotonically attracting trapping region
is disproved for the 8-dimensional system
which has a locally stable limit cycle 
but also solutions converging to infinity.
Thus, the theorems of section~\ref{sec:stabcrit}
proof numerically suggested behaviour
of the hierarchy of Galerkin models.

\subsection{On the long-term boundedness of a mean-field system}
\label{subsec:meanfieldsystem}

We consider a mean-field system 
for a soft onset of oscillatory fluctuations \citep{Noack2003jfm}
in fluid flows.
The state contains 3 coordinates:
$x_1$ and $x_2$ describe the amplitude of the phase of the oscillatory fluctuation
and $x_3$ characterises the mean-field deformation.
The origin $x_1=x_2=x_3=0$ corresponds to the steady solution.
For simplicity, 
many parameters of the general mean-field model 
\citep[see, e.g.,][]{Noack2011book} are set to zero or unity,
following Sec.\ 2.1 of \citet{Noack2003jfm}.
Only the bifurcation parameter $\mu$ is left.
The resulting mean-field system reads
\begin{subequations}
\label{eq:meanfieldmodel}
\begin{eqnarray}
\frac{dx_1}{dt} & = & \mu \, x_1 \; - \; x_2 \; - \; x_1 \, x_3\, ,\\
\frac{dx_2}{dt} & = & \mu \, x_2 \; + \; x_1 \; - \; x_2 \, x_3\, ,\\
\frac{dx_3}{dt} & = & - x_3 \; + \; x_1^2 \; + \; x_2^2.
\end{eqnarray}
\end{subequations}
and has the form of system~\eqref{eq:gspre}
with an energy-preserving quadratic term.
For subcritical Reynolds numbers ($\mu <0$), the system
has a globally stable fixed point $x_1=x_2=x_3=0$.
For supercritical values, ($\mu>0$) these
fixed points becomes unstable and all trajectories
converge to the limit cycle
\begin{eqnarray}
x_1 \; = \; \sqrt{\mu} \, \cos(t), \; x_2 \; = \; \sqrt{\mu} \, \sin(t), \; x_3 \; = \; \mu\, .
\end{eqnarray}
modulo an irrelevant phase.
This limit cycle represents vortex shedding.

This differential equation system can be brought in the  form~\eqref{eq:gs} with
\begin{eqnarray}
{\bfc}\; = \; \left[ \begin{array}{c} 0\\ 0\\ 0 \end{array}\right]\, ,
&& {\bfL}
\; = \; \left[ \begin{array}{ccc} \mu & -1 & 0\\ 1 & \mu & 0\\ 0 & 0 & -1  \end{array}\right]
\end{eqnarray}
and
\begin{eqnarray}
{\bfQ^{(1)}}
\; = \; \left[ \begin{array}{ccc} 0 & 0 & - \frac 1 2\\ 0 & 0 & 0\\ -\frac 1 2 & 0 & 0 \end{array}\right], \;
{\bfQ^{(2)}}
\; = \; \left[ \begin{array}{ccc} 0 & 0 & 0\\ 0 & 0 & - \frac 1 2\\ 0 & - \frac 1 2 & 0 \end{array}\right], \;
{\bfQ^{(3)}}
\; = \; \left[ \begin{array}{ccc} 1 & 0 & 0\\ 0 & 1 & 0\\ 0 & 0 & 0 \end{array}\right], \; \nonumber
\end{eqnarray}
with a real parameter $\mu>0$.

It may be interesting to note that 
system~\eqref{eq:simplesystem} 
originated from the mean-field model \eqref{eq:meanfieldmodel} with $\mu=1$.
The component~$x_1$ in~\eqref{eq:simplesystem} 
corresponds to the component~$x_3$ in~\eqref{eq:meanfieldmodel},
the state~$x_2$ in~\eqref{eq:simplesystem} 
to the amplitude $\sqrt{x_1^2+x_2^2}$ of the
two first components in~\eqref{eq:meanfieldmodel}.
Like in the example system~\eqref{eq:simplesystem}, 
there are areas in which the amplidutde of the oscillation
in the $x_1$-$x_2$-plane grows 
while the $x_3$-direction has only negative distance growth. 
The stabilisation of the  limit cycle can be described by the Landau equation~\citep{Noack2003jfm}.
Like for system~\eqref{eq:simplesystem}, 
the boundedness of the mean-field dynamics
cannot be derived from a quadratic Lyapunov function.

\begin{figure}
\begin{center}
\includegraphics[height=6.2cm]{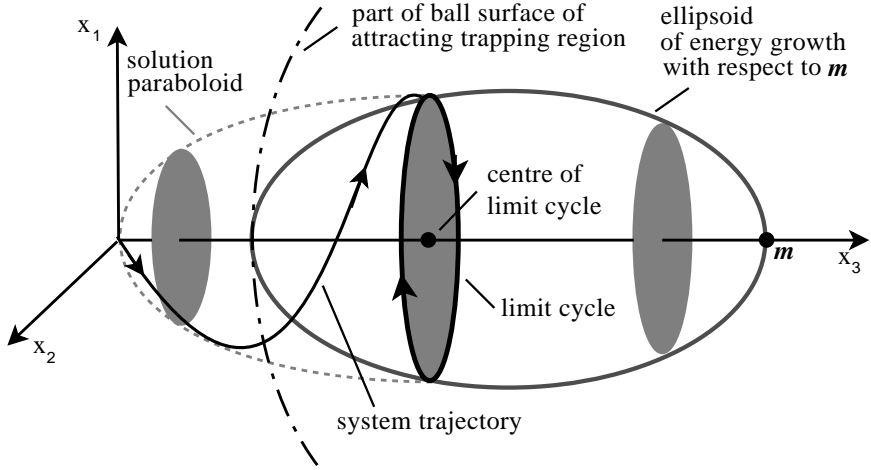}
\caption{Solution behaviour of the system~\eqref{eq:meanfieldmodel}. The trajectories
converge in spirals along the paraboloid (light grey, dashed line)
toward the globally stable limit cycle (black, solid and thick line),
as exemplified by one representative (black, solid, thin line).
For one $\bfm$ with a corresponding negative definite $\bfA_S$ the trapping region of minimal
radius (indicated by black, dot-dashed line) is determined by the contained
ellipsoid of positive energy growth~\eqref{eq:ellipsoid} (dark grey, solid, thick line).
}
\label{fig:paraboloid}
\end{center}
\end{figure}

The criterion of theorem~\ref{theorem1} can easily be satisfied.
Consider the translation with~${\bfm}=[0,0,\mu+\epsilon]^\top$ 
for an ~$\epsilon>0$.
Employing moreover~\eqref{eq:symmlin}, 
the symmetric part of the transformed system is derived to be
\begin{eqnarray}
{\bfA}_S & = &
\left[ \begin{array}{ccc} -\epsilon & 0 & 0\\ 0 & -\epsilon & 0\\ 0 & 0 & -1  \end{array}\right].
\nonumber
\end{eqnarray}
The negative definiteness of~${\bfA}_S$ implies
the existence of a monotonically attracting trapping region.
Following theorem~\ref{theorem1}, 
one of these regions is given by the closed ball $B({\bfm},R_m)$,
where $R_m=(\mu+\epsilon)/\sqrt{\epsilon}$. 
Hence, the trapping region will grow to infinity for~$\epsilon\to 0$. 
The limit cycle is situated inside of $B$ 
and at the boundary of the ellipsoid~$E$ 
defined by equation~\eqref{eq:ellipsoid},
along which the energy~$K_{\bfm}$ is constant. 
This is illustrated in figure~\ref{fig:paraboloid}.
We employ that ${\bfm}$ lies on the $x_3$-axis orthogonal to the limit cycle plane, 
 $K_{\bfm}$ is seen to remain constant on the limit cycle.
The limit cycle is contained in the intersection set of
the infinite number of ellipsoids, 
each defined by a positive parameter $\epsilon>0$.

\subsection{On the long-term boundedness of a POD Galerkin model}
\label{subsec:8modemodel}

In \citet{Deane1991pfa} and \citet{Noack2003jfm}, 
a Galerkin model for the cylinder wake flow is proposed
employing the first 8 modes from a proper orthogonal decomposition~(POD).
The post-transient dynamics of oscillatory laminar vortex shedding at a Reynolds number
of $\Rey=100$ in the wake is accurately resolved by this model.

The 8-dimensional Galerkin system is given by the differential equation
\begin{eqnarray}
\frac{dx_i}{dt} & = & c_i + \sum_{j=1}^8 l_{ij} \, x_j 
 + \sum_{j,k=1}^8 q_{ijk} \, x_j x_k.
\label{eq:8modemodel}
\end{eqnarray}
The energy preservation property~\eqref{eq:enerypreservation} 
is enforced for the quadratic term.
The kinetic energy~$K$ is produced in the first mode pair ($x_1, x_2$)
with the same positive growth rate of energy.
The other six modes form pairs of the same negative energy growth rate,
consuming the energy transferred by the first pair.
For initial conditions close to the projection of the
Navier-Stokes attractor onto the 8-dimensional subspace, 
the post-transient dynamics is reproduced~\citep{Noack2003jfm}.
Starting far from the limit cycle, 
also solutions which converge to infinity
are numerically observed.

To test the criterion of theorem~\ref{theorem1}, 
the largest eigenvalue of the linear symmetric
part~${\bfA}_S$ is minimised over a set of shift vectors~${\bfm}$.
The optimisation has been performed via a simulated annealing
algorithm~\citep[see,~e.g.,][]{Gershenfeld2006book}
with random seeding in~${\bfm}\in[-100,100]^8$. 
This box contains 
the limit cycle 
which is centred around the origin.
The box can be considered as very large,
since it is almost two orders of magnitude larger 
then the radius of this limit cycle.
In all computations, 
the minimum of the largest eigenvalues is positive, 
which indicates that the criterion cannot be fulfilled. 
This result is  confirmed by simulations shown 
in~\citet{Noack2003jfm} 
verifying a divergent behaviour 
of~\eqref{eq:8modemodel} to infinity for some initial values.
Also \citet{Deane1991pfa} report fragile Galerkin system behaviour
for a similar POD wake model.
In fact, the fragility of the POD model for vortex shedding
has inspired numerous enhancements of the reduced-order modelling method,
e.g.\ 
a nonlinear eddy viscosity term \citep{Cordier2013ef},
a stabilising spectral viscosity term \citep{Sirisup2004jcp},
a stabilising linear term \citep{Galletti2004jfm},
a stabilising additional shift mode \citep{Noack2003jfm},
or the inclusion of Navier-Stokes constraints construction
of generalised POD modes \citep{Balajewicz2013jfm}.
 
\subsection{On the long-term boundedness of a POD Galerkin model with shift mode}
\label{subsec:2Dwakewithshiftmode}

The 8-dimension POD model of the previous section
is stabilised by including an additional shift mode~${\bfu}_9$ 
in the Galerkin expansion
following \citet{Noack2003jfm,Noack2005jfm}. 
This shift mode~${\bfu}_9$ 
represents the normalised difference of mean flow and stationary solution.
In addition, the base flow ${\bfu}_0$ 
is chosen to be the unstable steady solution 
so that the origin is the fixed point of the Navier-Stokes dynamics.
In the following, long-term boundedness of the
resulting 9-mode Galerkin system is proven for all initial conditions.

The dynamical system~\eqref{eq:8modemodel} 
is generalised by additional
terms on the right-hand sides 
and a new additional equation arising from the shift mode:
\begin{subequations}
\label{eq:9modemodel}
\begin{eqnarray}
\frac{dx_i}{dt} & = & \underbrace{c_i + \sum_{j=1}^8 l_{ij} \, x_j + \sum_{j,k=1}^8 q_{ijk} \, x_j x_k}_{
\mbox{terms from the 8-mode system~\eqref{eq:8modemodel}}} + \nonumber \\
&&\underbrace{l_{i9} \, x_9
+ \sum_{k=1}^8 q_{i9k} \, x_9 x_k
+ \sum_{j=1}^8 q_{ij9} \, x_j x_9 + q_{i99} \, x_9^2}_{\mbox{additional `shift mode' terms}}
\, , \; i=1,\ldots,8\, , \\
\frac{dx_9}{dt} & = & c_9 + \sum_{j=1}^9 l_{9j} \, x_j + \sum_{j,k=1}^9 q_{9jk} \, x_j x_k \, .
\end{eqnarray}
\end{subequations}
For all numerically investigated initial conditions 
employed in~\citet{Noack2003jfm},
the system solutions are long-term bounded and converge to a limit cycle.

\begin{figure}
\begin{center}
\includegraphics[height=4.7cm]{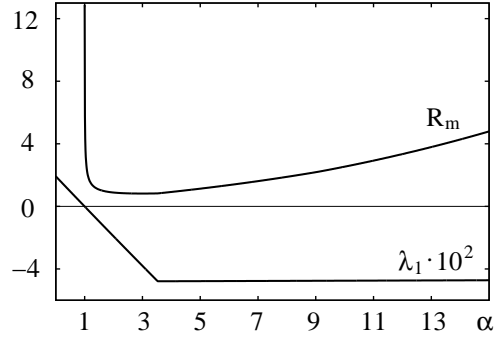}
\end{center}
\caption{Visualisation of the 
criterion of theorem~\ref{theorem1} for long-term boundedness of the 9-mode cylinder wake
Galerkin system~\eqref{eq:9modemodel}.
Shown are the largest eigenvalue $\lambda_1$ of ${\bfA}_S$ of the shifted system~\eqref{eq:gsshift}
and the estimated radius~$R_m$ of
the globally attracting trapping ball $B$ described in this theorem, for several shift
vectors~${\bfm}=\left(0,0,0,0,0,0,0,0,\alpha \right)^\top$ with $\alpha > 0$.}
\label{fig:cylsimulation}
\end{figure}

In the following, 
we prove long-term boundedness of the
9-mode Galerkin system with the criterion of theorem~\ref{theorem1}.
Learning from the mean-field model,
only translations~\eqref{eq:coordinateshift}
along the mean-field axis are considered,
i.e.\ ${\bfm}=\left(0,0,0,0,0,0,0,0,\alpha \right)^\top$ 
with $\alpha > 0$.
Figure~\ref{fig:cylsimulation} visualises the situation at $\alpha\approx1$. 
There is a change of the sign of the
largest eigenvalue of~${\bfA}_S$ from being positive to negative at $\alpha\approx 1$.
By the translation, 
the largest eigenvalue decreases initially linearly with~$\alpha$.
After some value of $\alpha$, this largest eigenvalue remains negative and constant
due other eigenvalues which are not affected by the translation.
Thus, the largest eigenvalue of~${\bfA}_S$ remains constant  for larger~$\alpha$.
In conclusion, there exist translations 
which make ${\bfA}_S$  negative definite.
Thus, the 9-mode Galerkin
system is shown to be long-term bounded 
because a monotonically attracting trapping region must exist 
according to theorem~\ref{theorem1}. 
Furthermore, it is
shown in the figure
that there is a minimum of the volume of the trapping region
for a certain ${\bfm}$ as indicated by the curve of the estimated radius $R_m$ of the trapping region.

\section{Design of a Trefethen-Reddy Galerkin system}
\label{sec:trefethenreddy}

In this section, 
we consider the Trefethen-Reddy system \citep{Trefethen1993science}
as celebrated paradigm for linear transient growth and nonlinear dynamics. 
This system has a non-algebraic nonlinearity, 
i.e.\ it cannot be obtained from a Galerkin method. 
Here, we use the criterion of theorem~\ref{theorem1} 
to design an energy-preserving quadratic term
with similar nonlinear behaviour.

\citet{Trefethen1993science} 
have proposed a simple dynamical system
exhibiting important features 
of the non-normal linear term and nonlinearity:
\begin{eqnarray}
\frac{d{\bfx}}{dt} & = &
\left[ \begin{array}{cc} -\Rey^{-1} & 1 \\ 0 & -2\, \Rey^{-1} \end{array} \right] \, {\bfx}
\, + \, \Vert{\bfx}\Vert \, \left[ \begin{array}{cc} 0 & -1 \\ 1 & 0 \end{array} \right] \, {\bfx} \, .
\label{eq:trefethenreddysystemoriginal}
\end{eqnarray}
Here, $\Rey$ mimics the effect of the  Reynolds number.
The linear part of this Trefethen-Reddy system 
is represented by a non-normal matrix
such that transient growth of the energy~$K$ 
can be obtained in the linear regime. 
The transient growth levels increase with $\Rey$.
At sufficiently large values, 
the nonlinear term becomes important 
and the following bootstrapping effect has been observed.
Initial conditions~${\bfx}(0)=[0,const.]^\top$ are numerically considered.
For small~$\Vert{\bfx}(0)\Vert$, 
the dynamics converge to the origin with linear growth rates.
For larger~$\Vert{\bfx}(0)\Vert$, 
the trajectories converge to fixed points~${\bfx}_S$,
$\Vert{\bfx}_S\Vert\approx 1$, 
with far larger growth rates than in the linear regime.

The Trefethen-Reddy system qualitatively displays 
important non-normal linear and nonlinear effects
observed, for instance, in boundary layers and internal flows.
However, \eqref{eq:trefethenreddysystemoriginal}
cannot be obtained from a Galerkin method.
There exist no spatial modes~${\bfu}_1$,~${\bfu}_2$ 
such that the projection onto the incompressible Navier-Stokes equation
yields  a square root expression like~$\Vert{\bfx}\Vert$.

The goal in this section 
is to model the bootstrapping effect 
by a long-term bounded Galerkin system of form~\eqref{eq:gspre}, 
thus offering an alternative 
to ~\eqref{eq:trefethenreddysystemoriginal}
which is arguably more physical.
Hence, a new nonlinear term needs to be identified.
The requested energy preservation of the quadratic term
strongly limits the number of free parameters from $10$ to $2$
\citep[see also][]{Bourgeois2013jfm}.
The corresponding dynamical system reads
\begin{eqnarray}
\frac{d{\bfx}}{dt} & = &
\left[ \begin{array}{cc} -\Rey^{-1} & 1 \\ 0 & -2\,\Rey^{-1} \end{array} \right] \, {\bfx}
\, + \, \delta \left( \sin\theta \left[ \begin{array}{c}  x_1 \, x_2 \\ -x_1^2 \end{array} \right] \, +
\, \cos\theta \left[ \begin{array}{c} -x_2^2 \\ x_1 \, x_2 \end{array} \right] \right)
\label{eq:trefethenreddysystem}
\end{eqnarray}
for some~$\delta>0$ and~$\theta\in(0,2\upi]$.
This system is of form~\eqref{eq:gs} with
\begin{eqnarray}
{\bfL}_S \; = \;
\left[ \begin{array}{cc} -\Rey^{-1} & 1/2 \\ 1/2 & -2\,\Rey^{-1} \end{array} \right]\, ,
\nonumber
\end{eqnarray}
and
\begin{eqnarray}
{\bfQ}^{(1)} \; = \; \delta
\left[ \begin{array}{cc} 0 & 2^{-1} \sin\theta\\
2^{-1}\sin\theta & -\cos\theta \end{array}\right], \;
{\bfQ}^{(2)} \; = \; \delta
\left[ \begin{array}{cc} -\sin\theta & 2^{-1}\cos\theta\\ 2^{-1}\cos\theta & 0 \end{array}\right].
\nonumber
\end{eqnarray}

The existence of a monotonically attracting trapping region
can be shown for the 
parameters~$\theta\approx \upi/2,\upi,3\upi/2,2\upi$
as follows.
For the translation vector
\begin{eqnarray}
{\bfm} & = & \delta^{-1}\, [\sin\theta,\cos\theta]^\top,
\nonumber
\end{eqnarray}
it yields 
\begin{eqnarray}
{\bfA}_S & = &
\left[ \begin{array}{cc} -\Rey^{-1}
+ 2^{-1} \sin2\theta& 0\\ 0 & -2\,\Rey^{-1} + 2^{-1} \sin2\theta\end{array}\right]
\; \approx \;
\left[ \begin{array}{cc} -\Rey^{-1} & 0\\ 0 & -2\,\Rey^{-1} \end{array}\right].
\nonumber
\end{eqnarray}
At large Reynolds number~$\Rey$, 
the same ${\bfm}$ can be employed to show the long-term boundedness
for $\theta \in [\upi/2,\upi]$ and $\theta \in [3\upi/2,2\upi]$.
Via a detailed analysis of ${\bfA}_S$ for arbitrary ${\bfm}$ 
(see the techniques
applied in the appendix section~\ref{sec:counterexamples}), 
it can be shown that the  criterion of theorem~\ref{theorem1}
cannot be fulfilled 
for sufficient large~$\Rey$ 
and $\theta \in (0,\upi/2)$ or $\theta \in (\upi,3\upi/2)$.
Hence, only a $\theta \in [\upi/2,\upi]$
or $\theta \in [3\upi/2,2\upi]$ 
represents a candidate to model flow attractor behaviour.

\begin{figure}
\begin{center}
\includegraphics[height=4.7cm]{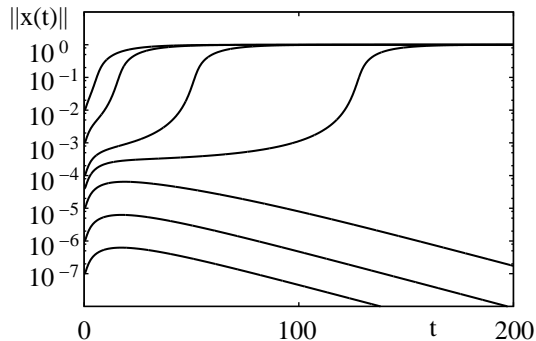}
\end{center}
\caption{Bootstrapping effect of the modified Trefethen-Reddy system. The evolution of
$\Vert{\bfx}(t)\Vert$ is visualised for solutions of~\eqref{eq:modifiedtrefethenreddysystem}
with initial conditions
$x_1(0)=0$, $x_2(0)=10^{-7},10^{-6},10^{-5},4\times10^{-5},10^{-4},10^{-3},10^{-2}$
and for $\delta =12.5$.}
\label{fig:trefsimulation}
\end{figure}

With the initial conditions of \citet{Trefethen1993science},
a similar bootstrapping effect can be
observed,  for instance for $\theta=3\upi/2$~(see figure~\ref{fig:trefsimulation}).
For this angle, 
the modified Trefethen-Reddy system 
reads
\begin{subequations}
\label{eq:modifiedtrefethenreddysystem}
\begin{eqnarray}
\frac{dx_1}{dt} & = & -\Rey^{-1} \, x_1 \; + x_2
\; - \; \delta \, x_1 \, x_2\, ,\\
\frac{dx_2}{dt} & = & -2\,\Rey^{-1} \, x_2
\; + \; \delta \, x_1^2\, .
\end{eqnarray}
\end{subequations}
The location of the fixed points scale with the parameter $\delta$. 
For a similar scaling
like in~\citet{Trefethen1993science}, 
$\delta=12.5$ is chosen for the computations illustrated
in figure~\ref{fig:trefsimulation}.

In conclusion, 
the interval of system parameters for $\theta$ is reduced
employing the criterion of theorem~\ref{theorem1}.
The dynamical behaviour of the modified system coincides 
qualitatively with the original Trefethen-Reddy system: 
the bootstrapping effect is modelled 
via a dynamical system
with the same linear term
but an energy-preserving quadratic term.
Contrary to the original model,
such a system may arise from 
a Galerkin projection of a 2-mode expansion 
onto the Navier-Stokes equation.

\section{Long-term boundedness of the Lorenz system}
\label{sec:lorenzsystem}

The existence of a monotonically attracting trapping region 
is demonstrated for Galerkin systems~\eqref{eq:gspre} 
with stable fixed point behaviour of the example system~\eqref{eq:simplesystem} 
and the stable periodic limit cycle dynamics of system~\eqref{eq:meanfieldmodel}.
In this section and appendix~\ref{sec:counterexamples},
more complex examples are considered.
We start with the well-known Lorenz system
\begin{subequations}
\label{eq:lorenz}
\begin{eqnarray}
\frac{dx_1}{dt} & = & -\sigma \, x_1 \; + \; \sigma \, x_2\, ,\\
\frac{dx_2}{dt} & = & \rho \, x_1 \; - \; x_2 \; - \; x_1 \, x_3\, ,\\
\frac{dx_3}{dt} & = & -\beta \, x_3 \; + \; x_1 \, x_2\, ,
\end{eqnarray}
\end{subequations}
for positive parameters $\sigma$, $\rho$ and $\beta$. 
This system is of form~\eqref{eq:gs} with
${\bfc}={\bfzero}$, ${\bfQ^{(1)}} = {\bszero}$ and
\begin{eqnarray}
{\bfL} \; = \;
\left[ \begin{array}{ccc} -\sigma & \sigma & 0\\ \rho & -1 & 0\\ 0 & 0 & -\beta 
\end{array}\right], \;
{\bfQ^{(2)}} \; = \;
\left[ \begin{array}{ccc} 0 & 0 & - \frac  1 2\\ 0 & 0 & 0\\ -\frac 1 2 & 0 & 0 
\end{array}\right], \;
{\bfQ^{(3)}} \; = \;
\left[ \begin{array}{ccc} 0 & \frac 1 2 & 0\\ \frac 1 2 & 0 & 0\\ 0 & 0 & 0 
\end{array}\right]\, .
\label{eq:lorenz2}
\end{eqnarray}
In particular, the quadratic term is energy preserving.
For Lorenz's choice of the parameters 
\begin{equation}
\label{eq:LorenzParameter}
\sigma=10, \quad \rho=28, \quad \beta=8/3
\end{equation}
the solution is characterised by a strange attractor.
However it is known, that the solution is long-term bounded.
A trapping region of ellipsoidal form can found via a
Lyapunov function~\citep{SwinnertonDyer2001physletta}.

\begin{figure}
\includegraphics[height=3cm]{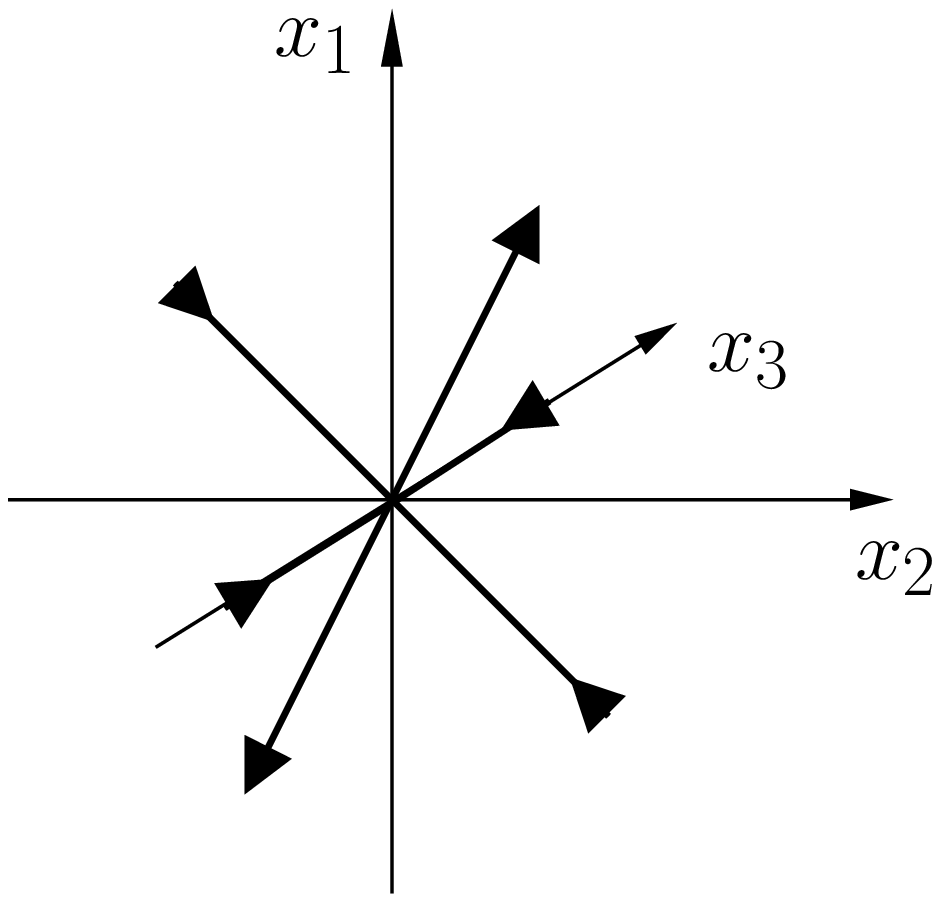}
\hspace{.1cm} \raisebox{1cm}{\Huge +} \hspace{.1cm}
\includegraphics[height=3cm]{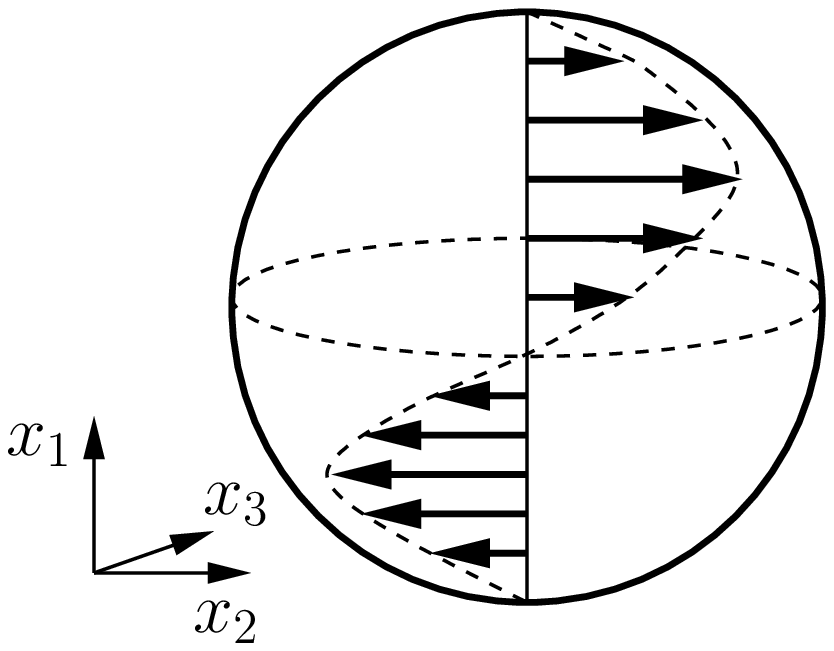}
\hspace{.1cm} \raisebox{1cm}{\Huge =} \hspace{.1cm}
\includegraphics[height=3cm]{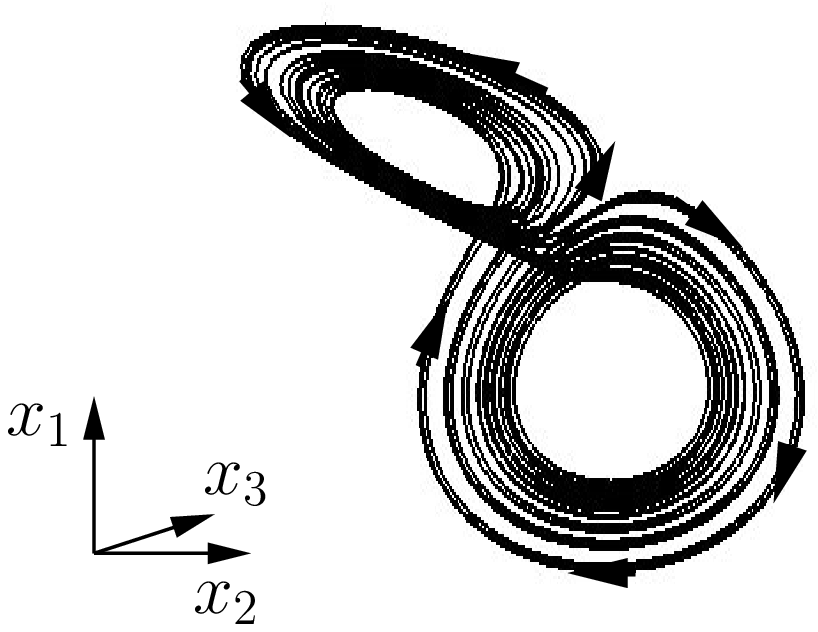}
\caption{Fields of the symmetric part of the linear term (left) and the quadratic
term (middle),
and the strange attractor (right) of the Lorenz system~\eqref{eq:lorenz} for Lorenz's
choice
of system parameters.}
\label{fig:3Ddynamics}
\end{figure}

Because there are positive and negative eigenvalues of ${\bfL}_S$, 
there exist directions of positive and
of negative energy growth $K$. 
However, via the quadratic term the trajectories are deflected from directions
of positive energy growth to directions of negative energy growth, stabilising the
resulting
strange attractor (see figure~\ref{fig:3Ddynamics}).

The directions of energy growth 
and the symmetry axes of the quadratic term do not coincide. 
The dynamics of the system is dominated by the quadratic term 
in case of large deviations from the origin 
and far enough from fixed points of the quadratic term
at the two poles and the equator. 
Hence, 
the boundedness of the system is determined by
the accumulation of energy growth along the trajectories of the quadratic term
crossing areas of negative and positive growth.
The set of points with a vanishing quadratic term 
has to be investigated separately,
leading to an investigation of the linear term.

In the following,  long-term boundedness is shown
with the criterion of theorem~\ref{theorem1}. 
After the translation
employing~${\bfm}=[0,0,\rho+\sigma]^\top$ and invoking~\eqref{eq:symmlin}, 
the symmetric part of the transformed system is equal to
\begin{eqnarray}
{\bfA}_S & = & 
\left[ \begin{array}{ccc} -\sigma & 0 & 0\\ 0 & -1 & 0\\ 0 & 0 & -\beta
\end{array}\right].
\nonumber
\end{eqnarray}
The negative definiteness of the matrix ${\bfA}_S$ proofs
the existence of a monotonically attracting trapping region.
Following theorem~\ref{theorem1} further, 
one of these regions is given by the closed ball $B({\bfm},R_m)$
with $R_m=\beta(\rho+\sigma)/\sqrt{\sigma}$ in case of $\sigma>\beta>1$.
For Lorenz's choice of parameters \eqref{eq:LorenzParameter},
the trapping region is given by $B({\bfm},R_m)\approx B([0,0,38]^\top,32)$.
Hence, the complete strange attractor is situated inside of the ball $B$.
It can be shown that the Lorenz system represents a prototype of the dynamics
illustrated in figure~\ref{fig:ellipsoid}
via further computation of the ellipsoid~$E$ using~\eqref{eq:ellipsoid}.
For large times the trajectories of the Lorenz system are pushing
the boundary of the ellipsoid through
and are alternately repelling from and attracting to~${\bfm}$.

\section{Conclusions and future directions}
\label{sec:conclusions}

We consider linear-quadratic dynamical systems
and propose a criterion
 which is sufficient for long-term boundedness 
and necessary for globally stable attractor behaviour.
For the first time, 
a straight-forward procedure~(see figure~\ref{fig:decisiontree})
can discriminate between physical and  unphysical behaviour
for a generic class of Galerkin models with quadratic nonlinearity.
The key enabler is  a generalisation 
of Lyapunov's direct method for identification
of monotonically attracting trapping regions.
These regions  represent a more accessible property
than the attractor property: 
the existence of monotonically attracting trapping regions
is based solely on eigenvalue computations of linear combinations of
system intrinsic matrices.

One distinct benefit is given for the model calibration and model reduction:
unphysical systems can be identified a priori.
One can avoid 
the computational burden of integration of the dynamical systems needed
for a comprehensive set of system parameters 
and a large set of initial conditions.

Similarly, control laws 
which lead to unphysical nonlinear dynamical behaviour can be rejected a priori.
A straight-forward control design is enabled via the design of monotonically attracting trapping regions. 
In the appendix section~\ref{sec:control}, 
the respective control design
is detailed for state stabilisation and for attractor control.

The criterion is applied 
to reduced-order models like the Galerkin models for a cylinder
wake~\citep{Noack2003jfm}, the Lorenz system,
and modifications to show long-term boundedness or indicate unboundedness (see table~\ref{table}).
Furthermore the capability of the criterion 
to reduce complexity for parameter identification is
demonstrated for the Trefethen-Reddy system. 
Here, a Galerkin system is identified, reproducing
the bootstrapping phenomenon observed in~\citet{Trefethen1993science}.
\begin{table}
\begin{center}
\begin{tabular}{@{}lclclcl@{}}
{Class 1 systems} && {Class 2 systems} && {Class 3 systems}\\
&& &&\\
9-mode Galerkin system~\eqref{eq:9modemodel} && 8-mode Galerkin system~\eqref{eq:9modemodel} && \\
for the cylinder wake, && for the cylinder wake && \\
mean-field system~\eqref{eq:meanfieldmodel} && && \\
&& && \\
Lorenz system~\eqref{eq:lorenz} && modification~\eqref{eq:Lorenzmodified1} of Lorenz &&
modification~\eqref{eq:Lorenzmodified1} of Lorenz \\
&& system for $\alpha_1 \geq 0$ &&  system for $\alpha_1 < 0$\\
&& && \\
'example system'~\eqref{eq:simplesystem} && 'inversed example system'~\eqref{eq:invsimplesystem}&&\\
&& && \\
modification~\eqref{eq:trefethenreddysystem} of &&
modification~\eqref{eq:trefethenreddysystem} of &&\\
the Trefethen-Reddy system && the Trefethen-Reddy system &&\\
for $\theta=0,\upi/2$ && for $\theta\neq0,\upi/2$&&\\
&& && \\
&& Rikitake system~\eqref{eq:rikitake} &&\\
&& && \\
&& Hamiltonian systems && \\[8pt]
\end{tabular}
\end{center}
\caption{Classification of the investigated dynamical systems
employing the categories introduced in figure~\ref{fig:decisiontree}.}
\label{table}
\end{table}

The proposed methods are generalisable 
to Galerkin systems of larger dimensions in a straight-forward manner.
Such systems may originate, for instance, from computational fluid dynamics. 
The numerical realisation of the criterion is only restricted 
by the current state of the art of
the numerical linear algebra and multidimensional optimisation.

\begin{acknowledgments}
\section*{Acknowledgements}

The authors acknowledge the funding and excellent working conditions
of the Chaire d'Excellence 'Closed-loop control of turbulent shear flows
using reduced-order models' (TUCOROM) of the French Angence Nationale
de la Recherche (ANR) hosted by Institute PPRIME.
We appreciate valuable stimulating discussions with
Markus Abel,
Laurent Cordier,
Eurika Kaiser,
Jean-Charles Laurentie,
Marek Morzy\'nski,
Robert Niven,
Vladimir Parezanovic,
J\"orn Sesterhenn,
Marc Segond,
and Andreas Spohn as well as the support by
the Deutsche Forschungsgemeinschaft (DFG)
under grants NO$\>$258/1-1, NO$\>$258/2-3, SCHL$\>$586/1-1
and SCHL$\>$586/2-1.
We are grateful for outstanding computer and software support
from Martin Franke and Lars Oergel.

\end{acknowledgments}



\appendix

\section{Boundary conditions for the preservation property of the quadratic term}
\label{sec:preservation}

In this section, 
property~\eqref{eq:enerypreservation} is derived for a large class of boundary conditions.
The energy preservation of the quadratic term it is known for periodic boundary
conditions~\citep[see, e.g.][]{McComb1991book,Holmes2012book} and 
for stationary Dirichlet conditions~\citep{Rummler2000hab}.
Here, we extend the range of validity also for open shear flows.

We assume an incompressible flow in a stationary domain $\Omega$ 
with Dirichlet, periodic or uniform free-stream conditions.
$\mathbf{\xi} \in \Omega$ represents the physical location.
The boundary of the domain is denoted by $\partial \Omega$.

Starting point are the quadratic Galerkin system coefficients
corresponding to the convective Navier-Stokes term
\citep[see, e.g., equation~(20) of the chapter `Galerkin method for Nonlinear Dynamics' in][]{Noack2011book}.: 
\begin{eqnarray}
\tilde{q}_{ijk} &:=& - \int_{\Omega} {\bfu}_i^{\top}
\, \bnabla\bcdot \left({\bfu}_j \, {\bfu}_k^{\top} \right) d{\bfxi}\, .
\label{eq:qtildeijk}
\end{eqnarray}

The symmetry~\eqref{eq:qijksymmetry}
of the Galerkin system coefficients~$q_{ijk}$ 
originated from the symmetrisation
\begin{eqnarray}
q_{ijk} = \frac 1 2 \left( \tilde{q}_{ijk} \; + \; \tilde{q}_{ikj} \right)\, .
\end{eqnarray}
It follows that equation~\eqref{eq:enerypreservation} is equivalent to
\begin{eqnarray}
\tilde{q}_{ijk} + \tilde{q}_{ikj} + \tilde{q}_{jik} + \tilde{q}_{jki} + \tilde{q}_{kij} + \tilde{q}_{kji}
= 0, & \; & i,j,k=1,\ldots,N.
\label{eq:enerypreservationtilde}
\end{eqnarray}

Equation~\eqref{eq:qtildeijk} is transformed by partial integration:
\begin{eqnarray}
\tilde{q}_{ijk}&=& - \int_{\Omega} {\bfu}_i \left(
\sum_{\beta=1}^3 u_j^\beta \frac{\partial}{\partial \xi_\beta} {\bfu}_k
\; + \; \underbrace{\left( \bnabla\bcdot {\bfu}_j \right)}_{=0} \, {\bfu}_k \right) d{\bfxi} \\
&=&
-\int_{\Omega} \sum_{\alpha,\beta=1}^3 \left(
u_i^{\alpha} u_j^\beta \frac{\partial}{\partial \xi_\beta} u_k^{\alpha} \right) d{\bfxi}\, , \nonumber\\
&=& - \int_{\Omega} \sum_{\alpha,\beta=1}^3 \left(
\frac{\partial}{\partial \xi_\beta} \left( u_i^{\alpha} u_j^\beta u_k^{\alpha} \right) \; - \;
u_k^\alpha \frac{\partial}{\partial \xi_\beta} \left( u_i^{\alpha} u_j^{\beta} \right) 
\right) d{\bfxi} \, ,\nonumber\\
&=& - \int_{\Omega} \bnabla\bcdot
\left( \left({\bfu}_i {\bfu}_k \right) {\bfu}_j\right) d{\bf\xi} \; + \;
\int_{\Omega} \left( {\bfu}_i {\bfu}_k \right)
\underbrace{\bnabla\bcdot {\bfu}_j}_{=0} d{\bfxi} \; + \; \int_{\Omega} 
\sum_{\alpha,\beta=1}^3 u_k^{\alpha} u_j^\beta \frac{\partial}{\partial \xi_\beta} u_i^{\alpha} d{\bfxi}
\, ,\nonumber\\
&=& - \oint_{\partial\Omega} \left({\bfu}_i {\bfu}_k \right) {\bfu}_j
{\bfeta} \, dS \; - \; \tilde{q}_{kji}\, ,
\label{eq:integralboundarycondition}
\end{eqnarray}
where~${\bfeta}$ denotes the unit outward normal at the surface~$\partial\Omega$ of the
considered domain~$\Omega$. Here, the
incompressibility of the modes~$\bnabla\bcdot {\bfu}_i=0$,~$i=1,\ldots,N$, is derived
from the postulated incompressibility of the flow~$\bnabla\bcdot {\bfu}$ 
exploiting the linearity of
the Galerkin approximation~\eqref{eq:galerkinapproximation} and of the divergence operator.

Let us assume vanishing surface integrals,
\begin{eqnarray}
\oint_{\partial\Omega} \left({\bfu}_i {\bfu}_j \right) {\bfu}_k = 0,
\quad i,j,k=1,\ldots,N.
\label{eq:integrals}
\end{eqnarray}
Then,~\eqref{eq:integralboundarycondition} and~\eqref{eq:integrals} imply
\begin{equation}
\label{eq:qijktildeSymmetry}
\tilde{q}_{ijk}=-\tilde{q}_{kji}, \quad
i,j,k=1,\ldots, N.
\end{equation}
The energy preservation of the original and symmetrised quadratic terms, 
i.e.\ \eqref{eq:enerypreservation} and \eqref{eq:enerypreservationtilde},
can easily be derived from \eqref{eq:qijktildeSymmetry}.

In the following, conditions for 
vanishing surface integrals~\eqref{eq:integrals} are determined.
\eqref{eq:integrals}
has been shown to vanish for stationary Dirichlet conditions,
implying the no-slip condition for all modes ~${\bfu}_i\equiv {\bfzero}$ at the boundary
\citep[see, e.g.][]{Rummler2000hab}.
Similarly straight-forward is the proof for periodic boundary conditions
\citep[see, e.g.][]{McComb1991book,Holmes2012book}.
Both boundary conditions may be combined, 
like in plane parallel Couette and Poiseuille channel flows~\citep{Rummler1998nnfm} or
in Hagen-Poiseuille flow~\citep{Boberg1988naturforschung}.

The proof for free-stream conditions at infinity is more challenging,
as it requires certain far-wake properties of the flow.
We consider flows in infinite domains around obstacles of finite extend. 
The integrals~\eqref{eq:integrals} vanish if the velocity rapidly decreases
for large distances from the origin. 
As examples of free turbulent shear flows, 
nominally two-dimensional cylinder wakes, 
two-dimensional jets 
and three-dimensional circular jets are considered. 
Here, it is easily shown 
that a sequence of integrals~\eqref{eq:integrals} over
surfaces of concentric balls~$\Omega_R$ converge to zero, 
if their radii~$R$ converge to infinity.
As centre of the balls~$\Omega_R$, 
the centre of the cylinder or the
centre of the nozzle exit might by employed. 
An upper bound of the absolute values of the integrals
is estimated via the radius, the dimension~$D=2,3$ 
of the flow configuration and the centre line (fluctuation) velocity $u_c$ in streamwise direction. 

The surface integral over the sphere $\Omega_R$ reads
\begin{eqnarray}
\left| \oint_{\partial\Omega_R} \left({\bfu}_i {\bfu}_j \right) {\bfu}_k
{\bfeta} \, dS \right| & \leq & 
\oint_{\partial\Omega_k} |{\bfu}_i| |{\bfu}_j| |{\bfu}_k| |{\bf\eta}| \, dS \, ,\nonumber\\
& \leq & \sup{|{\bfu}_i|} \, \sup{|{\bfu}_j|} \, \sup{|{\bfu}_k|} |\Omega_R|\, ,\nonumber\\
& \propto & u_c(R)^3 \, R^{D-1}\,.
\label{eq:intestimate}
\end{eqnarray}
Following~\citet{Schlichting1968book}, 
the centre line velocity is decreasing with radius~$R$ via
\begin{eqnarray}
u_c(R) \propto R^{\kappa}
\end{eqnarray}
where the decay rate~$\kappa$ is equal to $-1/2$ for two-dimensional wakes and jets,
and $\kappa=-1$ for the circular jet. 
Employing the estimate~\eqref{eq:intestimate}, 
in summary, there is a decrease of the absolute value of the integrals~\eqref{eq:integrals}
with $R^{-1/2}$ for two-dimensional wakes and jets and with $R^{-1}$ for the three-dimensional
circular jet. 
Hence, the integrals converge to zero for $R\to\infty$ 
leading to vanishing integrals over the infinite domain.

In conclusion, an intrinsic preservation property~\eqref{eq:enerypreservation} exists for
Galerkin systems of configurations with corresponding boundary conditions, 
e.g.\ flows around obstacles like spheres, cylinders and airfoils, in uniform stream as
well as many internal flows.

\section{Examples of unboundedness, vanishing nonlinearity and semidefiniteness}
\label{sec:counterexamples}

In this section, dynamical systems \eqref{eq:gspre} 
are discussed 
which do not obey the criterion for boundedness of theorem~\ref{theorem1}.
The examples include a modification of the two-dimensional system~\eqref{eq:simplesystem}
(subsection~\ref{sec:counterexample:2DSODE}),
modified Lorenz equations (subsection~\ref{sec:counterexample:Lorenz}),
and the Rikitake system (subsection~\ref{sec:counterexample:Rikitake}).

\subsection{Two-dimensional system of ordinary differential equations}
\label{sec:counterexample:2DSODE}
A first example for divergent dynamical behaviour is given 
from the time inversion $t \mapsto -t$
of~\eqref{eq:simplesystem}.
The resulting system has propagators of opposite sign and reads
\begin{subequations}
\label{eq:invsimplesystem}
\begin{eqnarray}
\frac{dx_1}{dt} & = & x_1 \; - \; x_2^2 \, ,  \\
\frac{dx_2}{dt} & = & -x_2 \; + \; x_1 \, x_2 \, ,
\end{eqnarray}
\end{subequations}
where
\begin{eqnarray}
{\bfc} \;=\; \left[ \begin{array}{c}  0\\ 0  \end{array}\right], \; 
{\bfL} \; = \; \left[ \begin{array}{cc} 1 & 0\\ 0 & -1 \end{array}\right], \;
{\bfQ^{(1)}} \; = \; \left[ \begin{array}{cc} 0 & 0 \\ 0 & -1  \end{array}\right], \;
{\bfQ^{(2)}} \; = \; \left[ \begin{array}{cc} 0 & \frac 1 2\\ \frac 1 2 & 0 \end{array}\right].
\label{eq:invsimplesystem2}
\end{eqnarray}
The long-term behaviour of system~\eqref{eq:invsimplesystem} is investigated 
employing the criterion of theorem~\ref{theorem1}.
Invoking~\eqref{eq:symmlin} and employing the matrices~\eqref{eq:invsimplesystem2},
the symmetric part of the shift transformed system is determined to be
\begin{eqnarray}
{\bfA}_S & = &
\left[ \begin{array}{cc} 1 & -{m_2}/2\\ -{m_2}/2 & m_1-1  \end{array}\right].\nonumber
\end{eqnarray}
The larger eigenvalue of ${\bfA}_S$
\begin{eqnarray}
\lambda_{1} & = & \frac 1 2 \left( m_1 + \sqrt{(2-m_1)^2 + m_2^2} \right)\nonumber
\end{eqnarray}
is positive: $\lambda_1 \geq 1$ is concluded from $|2-m_1|\geq 2-m_1$ because
\begin{eqnarray}
m_1 + \sqrt{(2-m_1)^2 + m_2^2} \geq m_1 + |2-m_1| \geq 2.
\end{eqnarray}
Hence, a monotonically attracting trapping region and a globally stable
attractor do not exist. 
By numerical system
integration, a divergent behaviour to infinity is observed for large times.

\subsection{Modified Lorenz system}
\label{sec:counterexample:Lorenz}
Similarly, the following modification of the Lorenz system
\begin{subequations}
\label{eq:Lorenzmodified1}
\begin{eqnarray}
\frac{dx_1}{dt} & = & \alpha_1 \, x_1\, ,\\
\frac{dx_2}{dt} & = & \alpha_2 \, x_2 \; - \; x_1 \, x_3\, ,\\
\frac{dx_3}{dt} & = & \alpha_3 \, x_3 \; + \; x_1 \, x_2\, ,
\end{eqnarray}
\end{subequations}
with $-\infty<\alpha_1,\alpha_2,\alpha_3<\infty$
is considered for investigation of existence of monotonically attracting trapping regions and
of globally stable attractors.
From equation~\eqref{eq:symmlin} we get
\begin{eqnarray}
{\bfA}_S & = & \left[
\begin{array}{ccc} \alpha_1 & -m_3/2 & m_2/2\\ -m_3/2 & \alpha_2 & 0\\ m_2/2 & 0 & \alpha_3 \end{array}
\right]. \nonumber
\end{eqnarray}
Independently of the choice of ${\bfm}$, 
the sum of the three eigenvalues of ${\bfA}_S$ is equal to the constant trace of ${\bfA}_S$,
i.e. the mean value
$\gamma = (\lambda_1 + \lambda_2 + \lambda_3 ) / 3$ is constant. 
The eigenvalues are increasingly separated 
with growing $|m_2|$, $|m_3|$ 
which can be seen from the dispersion of the eigenvalues
\begin{eqnarray}
\frac 1 3 \sum\limits_{i=1}^3 (\lambda_i - \gamma)^2 \; = \; 
\frac 1 3 \left( \sum\limits_{i=1}^3 \lambda_i^2 \right) - \gamma^2 & = &
f(\alpha_1,\alpha_2,\alpha_3) + \frac{m_2^2+m_3^2}6.
\label{eq:dispersionEigenValues}
\end{eqnarray}
For derivation of \eqref{eq:dispersionEigenValues}, 
Vieta's formula for the characteristic polynomial of ${\bfA}_S$
\begin{eqnarray}
(\lambda - \alpha_1)\,(\lambda - \alpha_2)\,(\lambda - \alpha_3)\;-\;
\frac{m_2^2}4\,(\lambda - \alpha_2)\;-\;
\frac{m_3^2}4\,(\lambda - \alpha_3) \nonumber
\end{eqnarray}
is employed, which yields
\begin{eqnarray}
\frac32 \gamma^2 - \frac12 (\lambda_1^2 + \lambda_2^2 + \lambda_3^2) \; = \;
\lambda_1 \lambda_2 + \lambda_2 \lambda_3 + \lambda_1 \lambda_3 \; = \;
\alpha_1 \alpha_2 + \alpha_2 \alpha_3 + \alpha_1 \alpha_3 - \frac{m_2^2+m_3^2}4\, . \nonumber
\end{eqnarray}
From the increased dispersion of the eigenvalues 
and the preservation of the mean value, 
the following can be concluded: 
If one of the three growth rates $\alpha_i$ is positive,
then for each ${\bfm}$ there is always at least one $\lambda_i$  positive as well.
Following theorem~\ref{theorem1},
a monotonically attracting trapping region does not exist in that case.
This result is easily validated considering the dynamics
in case that one eigenvalue is positive.
For $\alpha_1>0$, the variable $x_1$ is diverging to infinity for large times.
For $\alpha_1<0$ and large times the quadratic term is small compared to the linear term
and the variable $x_i$ with positive $\alpha_i$ is diverging to infinity.
If $\alpha_1=0$, $x_1$ is constant in time for each initial value,
i.e. a monotonically trapping region does not exist either.

In the case $\alpha_1<0$, the nonlinear term of the system~\eqref{eq:Lorenzmodified1} is vanishing
for large times. Invoking theorem~\ref{theorem2}, under action of an additional antisymmetric part,
there might exist globally attracting trapping regions or fixed points which are not monotonically attracting.
One example with an additional antisymmetric linear term is
\begin{subequations}
\begin{eqnarray}
\frac{dx_1}{dt} & = & \alpha_1 \, x_1,\\
\frac{dx_2}{dt} & = & \alpha_2 \, x_2 \; - \; x_3 \; - \; x_1 \, x_3,\\
\frac{dx_3}{dt} & = & \alpha_3 \, x_3 \; + \; x_2 \; + \; x_1 \, x_2.
\end{eqnarray}
\end{subequations}
The analysis for a monotonically attracting region is de facto the same
like in system~\eqref{eq:Lorenzmodified1}. 
However, if $\alpha_1<0$, the application of linear theory is enabled
by the vanishing of the nonlinear term for large times. 
Only the evolution equations for $x_2$ and $x_3$ are
considered  while the nonlinear term is neglected.
Here, $[x_2,x_3]^\top=[0,0]^\top$ is a globally stable fixed point,
if and only if the real parts of the eigenvalues of
$\left[ \begin{array}{cc} \alpha_2 & -1\\  1 & \alpha_3\end{array} \right]$ are negative-valued.
In conclusion, in case of a vanishing nonlinearity, 
linear analyses are needed in addition
for a complete analysis of the long-term boundedness of the system.

\subsection{The Rikitake system}
\label{sec:counterexample:Rikitake}
The criterion of theorem~\ref{theorem1}
cannot  be generalised from negative definite matrices~${\bfA}_S$
to negative semidefinite matrices~${\bfA}_S$ in a straight-forward manner. 
As a corresponding example, a modified Rikitake system
\begin{subequations}
\label{eq:rikitake}
\begin{eqnarray}
\frac{dx_1}{dt} & = & -\nu \, x_1 \; + \; x_2 \, x_3\, ,\\
\frac{dx_2}{dt} & = & -\alpha \, x_1 \; - \; \nu \, x_2 \; + \; x_1 \, x_3\, , \\
\frac{dx_3}{dt} & = & \theta \; - \omega \; x_1 \, x_2\, ,
\end{eqnarray}
\end{subequations}
is utilised with the positive parameters~$\nu$ and~$\alpha$~\citep{Goriely1998physicad}.
In the traditional Rikitake system, $\theta=1$ and $\omega=1$ are chosen.
Here $\omega$ is set to $\omega=2$ to fulfil the postulation~\eqref{eq:qijksymmetry}.
From similar analyses like above, the positive semidefiniteness of the largest eigenvalue,
i.e.~$\lambda_1\geq 0$, of~${\bfA}_S$ can be shown. In the following, ${\bfm}=[0,0,\alpha/2]^\top$
is considered, where~$\lambda_1=0$ and~$\lambda_2=\lambda_3=-\nu$ because
\begin{eqnarray}
{\bfA}_S & = & \left[
\begin{array}{ccc} -\nu & 0 & 0\\ 0 & -\nu & 0\\ 0 & 0 & 0 \end{array}
\right]. \nonumber
\end{eqnarray}
We leave $\theta=1$ at first. Here, it is known that
there is a simple solution ${\bfx}=\left[0,0,t\right]^\top$
which is diverging to infinity for larges times. 
However for $\theta=0$, long-term boundedness is
immediately ensured invoking the evolution equation~\eqref{eq:energyshift} 
for the chosen~${\bfA}_S$
and ${\bfd}={\bfzero}$: the energy $K_{\bfm}$ cannot grow in any direction. 
This means that
for negative semidefinite~${\bfA}_S$ additional information, e.g.~of the constant term might
be crucial for long-term boundedness analyses.

\section{Large-deviation and attractor control}
\label{sec:control}

\subsection{General considerations}

In this section, 
applications of the criterion of section~\ref{sec:stabcrit}
for control design are sketched. 

For control, the term~${\bfB}\, {\bfb}(t)$
is added to the right side of the Galerkin system~\eqref{eq:gs} leading to
\begin{eqnarray}
\frac{d{\bfx}}{dt} & = &
{\bfc} + {\bfL}\,{\bfx} + {\bfB}\,{\bfx} + 
\left[ \, {\bfx}^\top \, {\bfQ}^{(1)} \, {\bfx} \, ,
\ldots, \, {\bfx}^\top \, {\bfQ}^{(N)} \, {\bfx} \, \right]^\top
\; + \; {\bfB}\,{\bfb} \,
\label{eq:gscontrolledpre}
\end{eqnarray}
with the input matrix~${\bfB}$ 
and the input vector~${\bfb}={\bfb}(t)$.
Let us assume full-state feedback
with constant, linear and quadratic terms
\begin{eqnarray}
{\bfb} & = & {\bfc}^b 
+ {\bfL}^b\,{\bfx} 
+ \left[ \, {\bfx}^\top \, {\bfQ}^{b(1)} \, {\bfx} \, ,
\ldots, \, {\bfx}^\top \, {\bfQ}^{b(N)} \, {\bfx} \, \right]^\top
\label{eq:gscontrollaw}
\end{eqnarray}
with the free coefficients in ${\bfc}^b $, ${\bfL}^b$ and ${\bfQ}^{b(i)}$.
The actuation term reads 
\begin{eqnarray}
{\bfB}\,{\bfb}(t)
=  {\bfc}^g + {\bfB}^g\,{\bfx} +
\left[ \, {\bfx}^\top \, {\bfQ}^{g(1)} \, {\bfx} \, ,
\ldots, \, {\bfx}^\top \, {\bfQ}^{g(N)} \, {\bfx} \, \right]^\top\, ,
\end{eqnarray}
introducing the feedback vector ${\bfc}^g$, 
the feedback matrix ${\bfL}^g$,
and the symmetric matrices ${\bfQ}^{g(i)}$ 
for the quadratic term of feedback.
We request that the control law \eqref{eq:gscontrollaw}
is chosen to respect the energy preservation property
\begin{eqnarray}
q^{g(i)}_{jk} + q^{g(j)}_{ik} + q^{g(k)}_{ij} & = & 0, \quad i,j,k=1,\ldots,N\, .
\end{eqnarray}

In summary, the actuated system reads  
\begin{eqnarray}
\frac{d{\bfx}}{dt} & = & {\bfc}^a  + {\bfL}^a \, {\bfx} + 
\left[ \, {\bfx}^\top \,  {\bfQ}^{a(1)}  \, {\bfx}
+  {\bfx}^\top \, {\bfQ}^{a(N)}  \, {\bfx} \,  \right]^\top,
\label{eq:gscontrolled}
\end{eqnarray}
where ${\bfc}^a = {\bfc} + {\bfc}^g$,
      ${\bfL}^a = {\bfL} + {\bfL}^g$, 
and ${\bfQ}^{a(i)} = {\bfQ}^{(i)} + {\bfQ}^{g(i)}$, $i=1,\ldots,N$. 
The controlled system is of form~\eqref{eq:gs} and thus
contained in the class of dynamical systems considered in this paper. 
That's why the long-term behaviour of the corresponding dynamics 
can be investigated by the criterion of theorem~\ref{theorem1}.
In control design, the choice of the parameters~${\bfc}^b$,~${\bfL}^b$ and~${\bfQ}^{b(i)}$,~$i=1,\ldots,N$,
is restricted by constraints implied by the input matrix~${\bfL}^g$. 

Two tasks are pursued here
\begin{enumerate}
\item Large-deviation control:
The purpose of this part is twofold. One goal is the modification of the parameters of
system~\eqref{eq:gs} such that the existence of a monotonically attracting trapping region is ensured.
On the other hand, artefacts like blow-ups in model-based control design are precluded a priori.
\item Attractor control: Target is the manipulation of statistical attractor moments. Thereto, in
this paper, tools are provided to design the volume
and the location of monotonically attracting trapping regions.
In one extreme case, on which is focused here, a globally attracting fixed point is designed, i.e.
the attractor mean is equal to a fixed point and the higher central moments are zero.
Focus of large deviation control is the identification or creation of
monotonically attracting trapping regions.
This can be achieved as described in section~\ref{sec:stabcrit}.
\end{enumerate}

As one extreme case of attractor control, we assume a constant actuation, 
i.e.~vanishing $\bfL^b$ and $\bfQ^{b(i)}$.
The feedback vector~${\bfc}^g$ may be chosen such that 
each state~${\bfm}\in {\mathcal S}$ is a globally attracting fixed point of the feedback system.
This implies 
\begin{eqnarray}
0 = \, {\bfc}^a_i + \sum_{j=1}^N  l_{ij} \, m_j -
\sum_{j,k=1}^N  q^{(i)}_{jk} \,  m_j m_k \, ,
&&i=1,\ldots,N\,.
\end{eqnarray}
Then, the energy~$K_{\bfm}$ represents a Lyapunov function,
because the linear symmetric part ${\bfA}_S$ is negative definite.
The choice of the feedback vector~${\bfc}^g$ might be restricted
such that a globally stable fixed point is not attainable
due to the constraints implied by the input matrix~${\bfL}^g$ 
of the flow control configuration. 
However even in this case,
the first and second attractor moments can be estimated from the location and the
volume of the trapping region.
Thus, the moments can be manipulated by the design of the ellipsoid of energy growth
given by equation~\eqref{eq:ellipsoid}.
The effort of a corresponding volume force actuation 
for this attractor control might be large.
A trade-off between the attractor scaling via choice of~${\bfc}^g$ and 
the above discussed design of~${\mathcal S}$ might be necessary for model-based flow control applications.

Examples for the application of large deviation and attractor control are discussed
in the following.

\subsection{Example for large deviation control}

\begin{figure}
\begin{center}
\includegraphics[height=4cm]{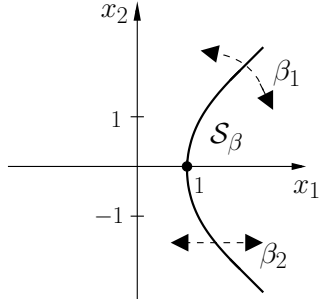}
\end{center}
\caption{Design of the open set~${\mathcal S}_{\bfbeta}$ of stabilisable states
at the left side of the curve.
By variation of the parameters~${\bfbeta}=[\beta_1,\beta_2]^\top$
of the feedback matrix~${\bfL}^g_{\bfbeta}$, the set~${\mathcal S}_{\bfbeta}$ is modified
via a widening/narrowing (varying~$\beta_1$) of the right opening angle
of the set or a spatial shift in $x_1$-direction (via varying~$\beta_2$).}
\label{fig:stableregion}
\end{figure}

For large deviation control, 
the set~${\mathcal S}$ of stabilisable states is analytically
or numerically created and designed
for the example of system~\eqref{eq:simplesystem} endowed with
a linear feedback matrix of the family of linear symmetric matrices
\begin{eqnarray}
{\bfL}^g_{\bfbeta} := \left[ \begin{array}{cc} \beta_1 & 0\\ 0 & \beta_2 \end{array} \right]\, .
\label{eq:controlmatrix}
\end{eqnarray}
The corresponding sets~${\mathcal S}_{\beta}$ of stabilisable states~${\bf m}$
are identified via auxiliary calculations to be
\begin{eqnarray}
{\mathcal S}_{\bfbeta} = \begin{cases}
\left\{ {\bfx} = [x_1,x_2]^\top : \; x_1>1+\beta_2, \, |x_2|<2\sqrt{1-\beta_1} \sqrt{x_1-1-\beta_2} \right\}
& \mbox{ for } \beta_1<1, \\
\emptyset & \mbox{ for } \beta_1\geq1 \, .
\end{cases}
\end{eqnarray}
In particular,
${\mathcal S}_{\bfzero} := \left\{ {\bfx} = [x_1,x_2]^\top : \; x_1>1, \, |x_2|<2\sqrt{x_1-1} \right\}$.
Starting from~${\mathcal S}_{\bfzero}$,
the set~${\mathcal S}_{\bfbeta}$ is designed via variation of~${\bfbeta}$
as demonstrated in figure~\ref{fig:stableregion}.
If the set~${\mathcal S}={\mathcal S}_{\bfbeta}$ is not empty,
the long-term dynamics of the resulting system is bounded invoking theorem~\ref{theorem1}.

In the case~$\beta_1 \geq 1$, the set of stabilisable states is empty and a
monotonically attracting trapping region does not exist, which is obvious considering the resulting system
\begin{eqnarray}
\frac{dx_1}{dt} & = & \left(\beta_1 - 1\right)x_1 \; + \; x_2^2\, ,  \\
\frac{dx_2}{dt} & = &  \left(1 + \beta_2\right) x_2 \; - \; x_1 \, x_2 \, ,
\end{eqnarray}
in the subspace of the $x_1$-axis, i.e.~$x_2=0$.
It means, that the~${\bfL}^g_{\bfbeta}$ with~$\beta_1\geq1$ cannot be chosen
to form a control with bounded system behaviour.

\subsection{Example for attractor control}

As attractor control example, we control the size of the monotonically attracting trapping region.
The Lorenz system~\eqref{eq:lorenz} is extended by a control 
vector~${\bfc}^g=\gamma\left[0,0,\beta(\rho+\sigma)\right]^\top$ with~$\gamma\in[0,1]$ such that
\begin{eqnarray}
\frac{d{\bfx}}{dt} & = & \gamma \left[ \begin{array}{c} 0\\ 0\\ \beta\left(\rho + \sigma\right) \end{array} \right] +
\left[ \begin{array}{ccc} -\sigma & \sigma & 0\\ -\rho & -1 & 0\\ 0 & 0 &-\beta \end{array} \right] {\bfx} +
 \left[ \begin{array}{c} 0\\ -x_1 \, x_3 \\ x_1 \, x_2 \end{array} \right] \, .
\end{eqnarray}
For~$\gamma=0$, this system is identical to the Lorenz equations~\eqref{eq:lorenz}.
From the transformation shift of~${\bfm}=[0,0,\rho+\sigma]^\top$
to the origin, the system for ${\bfy} = {\bfx} - {\bfm}$ is obtained:
\begin{eqnarray}
\frac{d{\bfy}}{dt} & = & \left( \gamma - 1 \right) \left[ \begin{array}{c} 0\\ 0\\ \beta\left(\rho + \sigma\right) \end{array} \right]  +
\left[ \begin{array}{ccc} -\sigma & \sigma & 0\\ -\sigma & -1 & 0\\ 0 & 0 &-\beta \end{array} \right]
{\bfy} +
\left[ \begin{array}{c} 0\\ -y_1 \, y_3 \\ y_1 \, y_2 \end{array} \right]
\end{eqnarray}
For~$\gamma=1$ the constant part of the right side is equal to zero.
Thus, the energy is a Lyapunov function and~${\bfm}$ is a globally stable fixed point in this case.
Generally, the equation~\eqref{eq:ellipsoid} for the ellipsoid of positive energy growth
is given in original coordinates by
\begin{eqnarray}
\frac{x_1^2}{\alpha_1^2} + \frac{x_2^2}{\alpha_2^2} + \frac{\left(x_3 - \left(\rho + \sigma\right) + \left(\gamma -1\right)\left(\rho + \sigma\right)/2\right)^2}{\alpha_3^2} & = & 1
\end{eqnarray}
with the half-axes
\begin{eqnarray}
\alpha_1\,=\,\left(1-\gamma\right)\frac{\rho+\sigma}{2\sqrt{\sigma}}, \;
\alpha_2\,=\,\left(1-\gamma\right)\frac{\rho+\sigma}{2}, \;
\alpha_3\,=\,\left(1-\gamma\right)\frac{\rho+\sigma}{2\sqrt{\beta}}\, .
\end{eqnarray}
Hence, the half-axes are shrinking linearly with the growth of~$\gamma$.
This is true as well as for the radius of the monotonically attracting trapping region
given by the smallest ball with ${\bfm}$ at the centre which contains the ellipsoid.

In conclusion, the attractor contained in this trapping region is shrinking,
and degenerates to a fixed point for~$\gamma=1$. The first statistical moment
situated in the ball is converging to~${\bfm}$ for~$\gamma \to 1$, the standard deviation
bounded by the ball radius is converging to zero.
For~$\gamma\approx 0.288$ there is a transition from the strange attractor
to three fixed points which converge for~$\gamma\to 1$ to the monotonically stable
fixed point ${\bfm}$ at $\gamma=1$. Thus, the control parameter $\gamma$ defines
a  transition scenario between
stationary and chaotic dynamical behaviour!

\end{document}